\documentclass[letterpaper, 9pt]{sigcomm-alternate}

\usepackage{color}
\usepackage{graphicx}
\usepackage{amsmath}
\usepackage{multirow}
\usepackage{subcaption}
\usepackage{enumitem}
\usepackage{comment}
\usepackage{color}
\usepackage{url}

\newcommand{\TODO}[1]{}
\renewcommand{\TODO}[1]{{\color{red} TODO: {#1}}}

\graphicspath{{./figures/}}

\begin{document}

\title{ISA-Based Trusted Network Functions And Server Applications In The Untrusted Cloud}

\numberofauthors{3}
\author{
\alignauthor
       Spyridon Mastorakis\\
       \affaddr{UCLA\thanks{This work is part of the author's internship at Intel Corporation.}}\\	
\alignauthor
       Tahrina Ahmed\\
       \affaddr{Intel Corporation}\\	
\alignauthor
       Jayaprakash Pisharath\\
       \affaddr{Intel Corporation}\\	  
}

\begin{comment}
% Double blind version
\author{
{\rm Author 1}\\
Affiliation 1
\and
{\rm Author 2}\\
Affiliation 2
\and
{\rm Author 3}\\
Affiliation 3
}
\end{comment}

\maketitle

\subsection*{Abstract}
Nowadays, enterprises widely deploy Network Functions (NFs) and server applications in the cloud. However, processing of sensitive data and trusted execution cannot be securely deployed in the untrusted cloud. Cloud providers themselves could accidentally leak private information (e.g., due to misconfigurations) or rogue users could exploit vulnerabilities of the providers' systems to compromise execution integrity, posing a threat to the confidentiality of internal enterprise and customer data.

In this paper, we identify (i) a number of NF and server application use-cases that trusted execution can be applied to, (ii) the assets and impact of compromising the private data and execution integrity of each use-case, and (iii) we leverage Intel's Software Guard Extensions (SGX) architecture to design Trusted Execution Environments (TEEs) for cloud-based NFs and server applications. We combine SGX with the Data Plane Development KIT (DPDK) to prototype and evaluate our TEEs for a number of application scenarios (Layer 2 frame and Layer 3 packet processing for plain and encrypted traffic, traffic load-balancing and backend server processing). Our results indicate that NFs involving plain traffic can achieve almost native performance (e.g., $\sim 22$ Million Packets Per Second for Layer 3 forwarding for 64-byte frames), while NFs involving encrypted traffic and server processing can still achieve competitive performance (e.g., $\sim 12$ Million Packets Per Second for server processing for 64-byte frames). 

\section {Introduction}
\label{intro}

Network Functions (NFs) deployed on routers, switches, and middleboxes are a vital part of today's network infrastructure, improving network performance, reliability, availability, and security. At the same time, large-scale applications require a large amount of storage and processing power. Both resources are vital for modern enterprises, which have to either deploy their own in-house IT infrastructure or use cloud-based network and server services. Previous work~\cite{sherry2012making} shows that the latter approach is less expensive and complicated in terms of management for enterprises, and more flexible and resilient in case of failures. As a result, nowadays, NFs and applications are widely deployed in the cloud and cloud routers, switches, middleboxes, and servers are commonly used by enterprises.

NFs and applications that maintain private data, process sensitive user data, and require trusted execution cannot be securely deployed in the untrusted cloud. In a cloud environment, providers may accidentally leak sensitive user information (e.g., because of a server misconfiguration) or malicious users may exploit vulnerabilities of the providers' systems~\cite{leak1, leak2, leak3} to compromise execution integrity. Such concerns about data confidentiality and execution integrity discourage enterprises from moving their entire operation to the cloud~\cite{santos2009towards}.

Previous work on securing NFs deployed in the cloud has considered to directly apply network processing over encrypted data~\cite{sherry2015blindbox, lan2016embark}. In this paper, we explore the approach of leveraging \emph{Intel's Software Guard Extensions (SGX)}~\cite{sgx1, sgx2} Instruction Set Architecture (ISA) to create Trusted Execution Environments (TEEs) for NF and server processing. To prototype our TEEs, we combine SGX with the \emph{Data Plane Development KIT (DPDK)}~\cite{dpdk}, which allows for rapid prototyping of high-performance data plane applications. Previous work on combining SGX and DPDK to provide a secure middlebox framework for cloud-based NFs~\cite{trach2017slick} does not consider packet switching, server processing, and NFs for the processing of encrypted traffic (e.g., VPN endpoints based on IPsec~\cite{kent2005ip} and MACsec capable switches~\cite{romanow2006media}). 

The limited study and experimentation with SGX-based TEEs for NFs and server processing motivated our work. We aim to contribute to the further understanding of the performance and trade-offs of applying SGX to cloud-based network and server solutions. Our contribution is twofold: (i) we discuss NF and server processing use-cases, where SGX can be applied to, and the assets and impact of compromising the private data and execution integrity of each use-case, and (ii) we present and evaluate proof-of-concept designs for a number of application scenarios (Layer 2 frame and Layer 3 packet processing for plain and encrypted traffic, traffic load-balancing, and backend server processing) through an experimental study.

The rest of our work is organized as follows: in section~\ref{background-related}, we give some background on the SGX and DPDK frameworks and discuss our related work. In section~\ref{threat-models}, we discuss NF and server application use-cases, where trusted execution can be applied to. In section~\ref{design}, we present our design approach. Section~\ref{implementation}, describes implementation-specific details of our work, and section~\ref{evaluation}, presents our experimental evaluation study. In section~\ref{discussion}, we describe the lessons learned and open issues of the current work that we plan to address in the future, and, finally, section~\ref{conclusions} concludes our work.
\section {Background \& Related Work}
\label{background-related}

In this section, we present a brief overview of SGX and DPDK to help the reader gain better understanding of what will be discussed in the rest of the paper. We also discuss some related work.

\subsection {Background}
\label{background}

\subsubsection{SGX}
\label{sgx-back}

The Intel SGX architecture offers a set of x86-64 ISA extensions that enable applications to instantiate a secure software container, called an~\emph{enclave}; an area in the virtual address space of the application, which is protected by the processor from accesses of any software that does not reside in it (e.g., other applications, OS, BIOS).

The enclave data is stored in a reserved memory cache, called Enclave Page Cache (EPC). The Memory Encryption Engine (MEE) encrypts the enclave data in EPC to avoid memory attacks (e.g., memory snooping). To access enclave data in EPC, the processor enters a new CPU mode, called \emph{enclave mode}, which applies additional hardware verifications to each memory access. Specifically, the data in EPC is decrypted only when entering the CPU package (enclave mode) and is encrypted again and stored to EPC when leaving the CPU package. 

Untrusted code can make incoming calls (ECALLs) to trusted enclave functions defined and exposed by developers, while enclave code can make outgoing calls (OCALLs) to untrusted code. In cases that the enclave execution is interrupted due to asynchronous events, such as interrupts and exceptions, the processor state is securely saved inside the enclave to prevent any leakage of secrets. After the event is serviced, the processor state can be restored and the enclave execution resumes from the point that was interrupted.

An enclave can prove that it has been properly instantiated on a platform through \emph{CPU-based attestation}. There are 2 attestation categories; \emph{local} and \emph{remote}. Local attestation enables 2 enclaves instantiated on the same platform to authenticate each other, while remote attestation enables an enclave instantiated on a remote platform to attest that it is ``trusted" to a remote attestation provider, so that secrets can be provisioned to it~\cite{johnson2016intel}. 

\subsubsection {DPDK}

DPDK consists of a set of libraries and optimized Network Interface Card (NIC) drivers for highly-scalable and fast packet processing, which is designed to run on any processor. It avoids the overhead imposed by Linux kernel processing (e.g., system calls, context switching on blocking I/O, copying data from kernel to user space, interrupts) and achieves high performance by: 1)  leveraging processor affinity, 2) allocating huge memory pages to avoid swaps and reduce TLB misses, 3) placing device drivers in user space to achieve zero-copy packet processing, 4) accessing all devices by polling, 5) achieving synchronization without locks, and 6) handling large batches of packets and distributing them to processing threads for unified processing. 

Each DPDK process (application) occupies one CPU core in full, but can actually use one or more of its logical cores. To exchange data among logical cores, lock-less First-In-First-Out (FIFO) ring structures are used. Each application can make use of DPDK libraries that provide network packet buffer management and packet forwarding mechanisms, and implement the TCP/IP protocol stack.

\subsection {Related Work}
\label{related-work}

In this section, we discuss software and hardware-based approaches that protect applications against unauthorized access. We also present some related work based on SGX and a few approaches that study the application of NFs directly to encrypted data.

\subsubsection {Software-Based Protection}

One of the very first works towards protecting applications and their sensitive data from unauthorized access by privileged software is~\emph{NGSCB}~\cite{peinado2004ngscb}. NGSCB made use of virtualization to run trusted and untrusted OSs simultaneously on the same machine enabling critical applications to use the trusted OS. A similar approach was also taken by~\emph{Proxos}~\cite{ta2006splitting} that requires application developers to specify which system calls are sensitive, so that they are forwarded to a trusted private OS, protecting applications against an untrusted OS.

Approaches, such as~\emph{Overshadow}~\cite{chen2008overshadow},~\emph{Virtual Ghost}~\cite{criswell2014virtual} and~\emph{InkTag}~\cite{hofmann2013inktag}, assumed a trusted virtualization layer to protect sensitive application data and aimed to reduce the size of TCB. Specifically, Overshadow offers different views of physical memory for each memory access, therefore, an application can have a normal view of its resources, but the OS an encrypted one. While Overshadow focuses on ensuring that applications are isolated from the OS, InkTag allows applications to use the services of an untrusted OS and define their own access control policies on secure files. Virtual Ghost utilizes compiler support to secure applications from an untrusted OS and creates secure memory, which cannot either be read or written by the OS.~\emph{MiniBox}~\cite{li2014minibox} is a two-way sandbox that protects critical applications from a malicious OS, as well as an OS from malicious applications.

All these approaches are purely based on software and do not require any special hardware support. Therefore, they can be used in cases that hardware-based solutions, such as SGX, cannot be deployed.

\subsubsection {Hardware-Based Protection}
\label{nets}

Several systems have utilized trusted hardware to secure applications running on them along with sensitive data from unauthorized access.

\emph{Trusted Platform Modules (TPMs)}~\cite{tpm} offer a dedicated micro-controller to offer secure generation of cryptographic keys and restrict their accessibility. TPMs also support remote attestation and data sealing functions. However, there are privacy concerns associated with the \emph{Direct Anonymous Attestation} (DAA) scheme used by TPM when a small number of keys is used for the entire platform lifetime~\cite{leung2008possible}. To address those concerns, SGX extends DAA by using an Enhanced Privacy ID (EPID) key during remote attestation~\cite{johnson2016intel}.

\emph{Secure co-processors}~\cite{smith1999building, dyer2001building} offer hardware that can be trusted even in cases of physical attacks, so that trusted computations can be performed on untrusted remote devices. However, they are expensive and their performance is limited due to thermal throttling issues.

\emph{ARM TrustZone}~\cite{arm2009security} is a security technology for System-on-a-Chip (SoC) and CPU systems based on the concept of physically separated trusted and untrusted worlds. It has been used to build an embedded virtualization system on commodity hardware~\cite{pinto2014towards} and a multi-layer security architecture for mobile devices~\cite{lengyel2014multi}. TrustZone is mainly used by embedded systems and does not offer memory encryption, therefore, attacks are possible in cases of physical DRAM access.

\emph{AMD's memory encryption technology}~\cite{kaplan2016amd} is integrated into the x86 CPU architecture and offers a security subsystem for key generation, platform boot, off-chip storage for sensitive data, protection against physical memory attacks and support for encrypted virtual machines. However, it specializes in memory encryption and does not provide a framework to run applications in trusted mode.

\subsubsection{SGX-Based Approaches}

\emph{SGX} offers CPU features that enable applications to instantiate secure and trusted enclaves. Research areas, to which SGX has been applied, include networked and distributed systems, cloud systems and applications. 

In Network Function Virtualization (NFV) environments, the design of an enclavized NAT, policy control and intrusion detection application, HTTP and web caching proxy~\cite{shih2016s}, and an extension to enclavize the Click modular router~\cite{coughlin2017trusted} have been presented. These approaches do not study the performance and overhead of applying SGX to frame processing and trusted encryption/decryption for IPsec and MACsec traffic, while the provided experimental results are limited.

Designs that explore how SGX could strengthen the security and privacy of peer-to-peer anonymity networks, such as Tor~\cite{kim2015first,kim2017enhancing}, network protocols, such as TLS~\cite{aublintalos}, and distributed services, such as the Apache ZooKeeper~\cite{brenner2016securekeeper}, have been studied. These protocols and services operate on a higher layer of the TCP/IP protocol stack and the performance of pure network forwarding and switching is not evaluated. Slick~\cite{trach2017slick} proposes a trusted middlebox framework to deploy network functions on untrusted servers. This work tackles a problem related to ours and some of their design decisions and optimizations can be used to further enhance our approach and vice versa.

Haven~\cite{baumann2015shielding} protects the confidentiality and integrity of applications and their associated data from the untrusted cloud on which they run, while VC3~\cite{schuster2015vc3} allows users to keep their data and secrets safe during the execution of distributed MapReduce computations in the cloud. The idea of an inverted cloud infrastructure has been discussed~\cite{strackx2015idea}, where mini providers use SGX to secure confidential information, so that they can join forces to provide cloud services instead of receiving services by a single major provider. SGX has also been used to secure content-based routing mechanisms~\cite{pires2016secure} and Database Management Systems (DBMS)~\cite{arasu2014querying} operating on the cloud. These pieces of work focus on trusted cloud applications and services running on top of today's networks, rather than the performance of the underlying network infrastructure itself.

\subsubsection{Network Functions Over Encrypted Data}

Starting with APLOMB~\cite{sherry2012making}, the idea of out-sourcing NF processing to the cloud emerged, without taking into account though its security considerations. BlindBox~\cite{sherry2015blindbox}, extending the approach taken by APLOMB, performs deep packet inspection directly over encrypted network traffic. Later on, Embark~\cite{lan2016embark} added support for a wider set of NFs over encrypted network data. This work focuses exclusively on protecting network traffic through encryption, rather than investigating how the execution of the NF software itself can be secured. Moreover, the studied middleboxes operate on the network layer and above, without  considering any link layer devices or server-side applications. 

\section {Trusted Execution For Network Functions and Server Applications}
\label{threat-models}

Each application and NF deployment processes is coupled with private data and executes software vital for its secure and legitimate operation. We present a few example use cases along with the assets of each case that can be protected through SGX in Table~\ref{table:use-cases} to show that trusted execution can apply to a wide set of systems and applications with different security concerns. We categorize the assets into \emph{data and data structures}, and \emph{software}. Given that SGX provides encrypted EPC to each enclave, whose access is forbidden to any untrusted entity, the data and data structures crucial for the operation of each use-case can be stored in EPC. Crucial software pieces can be executed in enclave mode, fully protected and isolated from the untrusted part of the system. 

To avoid exposing topology information, routing and management policies to attackers, a router should: 1) protect its internal data structures (e.g., routing table, Domain Name Server (DNS) cache, access control lists) by storing them in EPC, and 2) protect the actual software that performs operations using those structures (e.g., longest or exact IP prefix match, DNS cache lookup) by executing it in enclave mode. To avoid exposing forwarding policy information, legacy switches can store their forwarding table in EPC and execute in enclave mode all the operations related to it, while Software Defined Networking (SDN) switches can protect their flow table and the related operations from attackers.

The same applies to servers; for instance, DNS servers can leak information about the mapping of domain names to IP addresses, therefore, their DNS cache should be stored in protected enclave memory, while software that performs operations (e.g., lookups, insertions, deletions, updates) over this cache should be executed in enclave mode. Similarly, in a cloud multi-tenant environment, a server hosting multiple Virtual Machines (VMs) can secure each tenant's application instance and data in an enclave. 

Middleboxes, such as load balancers and firewalls can reveal to attackers load balancing policies and information about the blocked and accepted traffic respectively. To this end, they can protect their policies in EPC and enforce them to incoming traffic in enclave mode.

Since the source and destination IP address of a packet is typically in plaintext, on-path eavesdroppers can easily identify its source and destination. In a LAN, any Network Interface Card (NIC) that is present on it can listen to all the frames transmitted by any other NIC regardless of their destination MAC address, identifying the source and destination MAC address of the frames. To provide end-to-end security directly at the network and link layer of TCP/IP, security extensions have been added to the legacy IP and MAC protocols, formulated as the IPsec and MACsec protocols respectively. VPN endpoints and MACsec-capable switches make use of those protocols to encrypt network packets and Ethernet frames respectively. Since traffic encryption and authentication keys can be leaked, they should be stored in EPC, while encryption and authentication operations should be executed in enclave mode.

We should note that the size of EPC is currently limited to 128MB across all the enclaves, therefore, the data to be stored in it should be carefully selected. Exceeding the EPC size results in EPC paging (a mechanism for secure paging to the unprotected memory supported by SGX), which imposes additional execution performance overhead. We discuss how the performance impact from EPC paging can be alleviated in section~\ref{open-issues}.

% \usepackage{multirow}
\begin{table*}[h!]
\centering
\caption{Example Use-Cases \& SGX Protection}
\label{table:use-cases}
\begin{tabular}{|c|c|c|c|ll}
\cline{1-4}
\multirow{2}{*}{\textbf{Use-case}} & \multicolumn{2}{c|}{\textbf{Assets}}                                                                                                                                                                     & \multirow{2}{*}{\textbf{\begin{tabular}[c]{@{}c@{}}Protection through \\ SGX\end{tabular}}}                                &  &  \\ \cline{2-3}
                                   & \textit{Data and Structures}                                                  & \textit{Software}                                                                                                         &                                                                                                                            &  &  \\ \cline{1-4}
\textit{Network Router}            & \begin{tabular}[c]{@{}c@{}}Routing table, DNS cache \\ Access control lists\end{tabular}         & \begin{tabular}[c]{@{}c@{}}Forwarding logic, longest prefix match, \\ DNS cache and routing table operations\end{tabular} & \multirow{9}{*}{\textit{\begin{tabular}[c]{@{}c@{}}Data stored in EPC\\ Software executed in\\ enclave mode\end{tabular}}} &  &  \\ \cline{1-3}
\textit{Network Switch}            & Forwarding table                                                             & Forwarding table operations                                                                                               &                                                                                                                            &  &  \\ \cline{1-3}
\textit{Firewall}                  & Policy rules                                                                 & Rule lookup and match                                                                                                     &                                                                                                                            &  &  \\ \cline{1-3}
\textit{Load Balancer}             & Load balancing rules                                                         & Rule lookup and match                                                                                                     &                                                                                                                            &  &  \\ \cline{1-3}
\textit{DNS Server}                & DNS cache                                                                    & DNS cache related operations                                                                                              &                                                                                                                            &  &  \\ \cline{1-3}
\textit{Server hosting VMs}        & Tenant's data                                                                & Tenant's application instances                                                                                            &                                                                                                                            &  &  \\ \cline{1-3}
\textit{VPN Endpoint (IPsec)}      & \begin{tabular}[c]{@{}c@{}}Encryption and \\ authentication keys\end{tabular}     & \begin{tabular}[c]{@{}c@{}}Encryption and \\ authentication operations\end{tabular}                                                                                                     &                                                                                                                            &  &  \\ \cline{1-3}
\textit{MACsec-Capable Switch}     & \begin{tabular}[c]{@{}c@{}}Encryption and \\ authentication keys, \\ forwarding table\end{tabular} & \begin{tabular}[c]{@{}c@{}}Encryption, authentication \\ and forwarding table operations\end{tabular}                                      &                                                                                                                            &  &  \\ \cline{1-3}
\textit{SDN-Capable Switch}        & Flow table                                                                   & Flow table operations, forwarding logic                                                                                   &                                                                                                                            &  &  \\ \cline{1-4}
\end{tabular}
\end{table*}

\section {Design}
\label{design}

Our design consists of 2 major building blocks: DPDK and SGX. We first present a baseline approach utilizing a DPDK application and a single SGX packet processing enclave as well as the work-flow facilitated by this approach. We also present an approach to scale trusted processing and an approach to implement a trusted processing pipeline by leveraging multiple local enclaves and the SGX local attestation feature. 

\subsection {Baseline Approach}
\label {baseline}

\begin{figure}[h]
\centering
\includegraphics[scale=0.55]{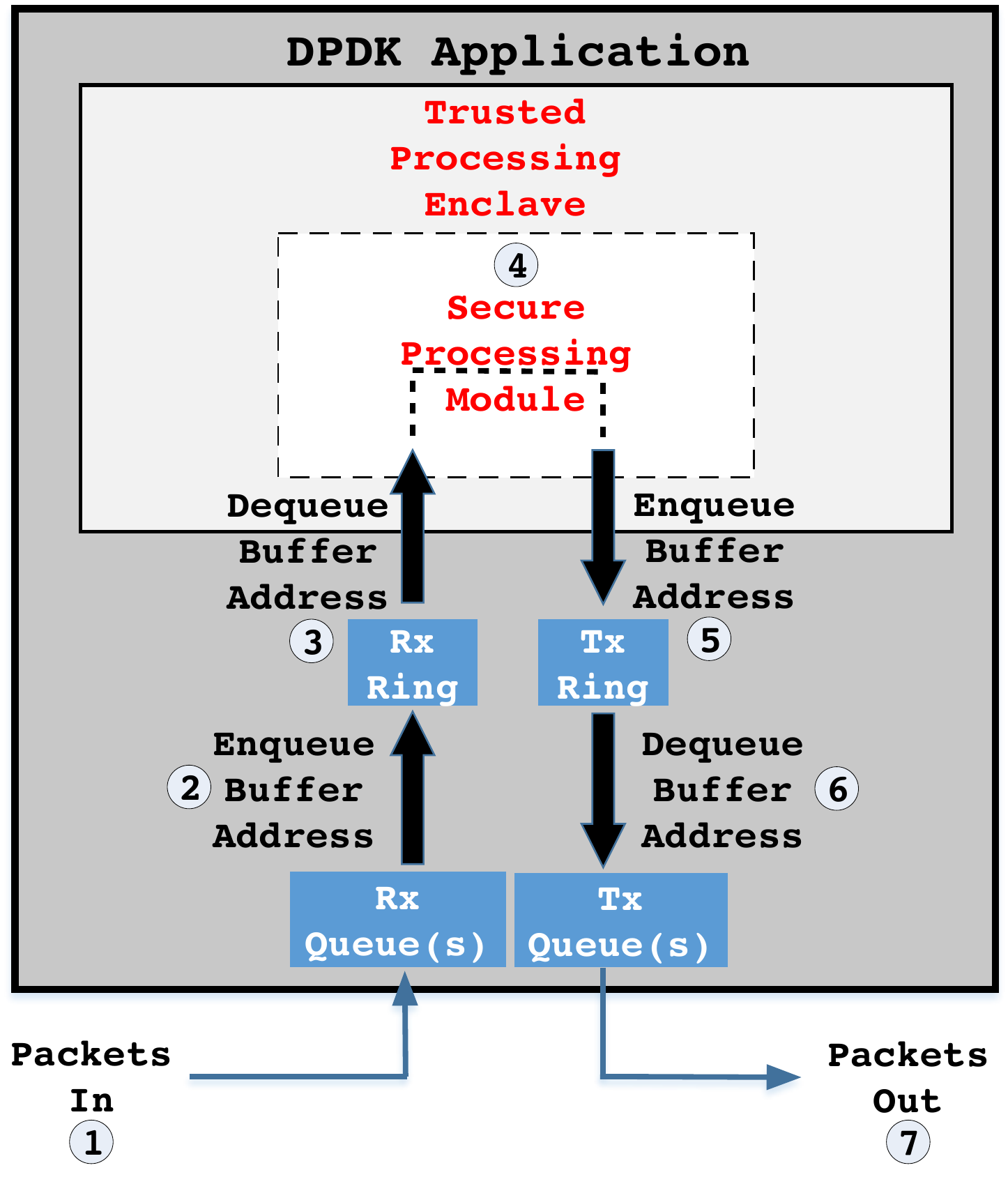}
\caption {Secure Packet Processing Design}
\label{Figure:design}
\end{figure}

Following the SGX application design principles~\cite{sgx1}, we divide our DPDK application into 2 parts; trusted, which is implemented within a processing enclave to protect vital data structures and code against unauthorized access, and untrusted, which refers to application code not protected by SGX. Our baseline design approach is illustrated in Figure~\ref{Figure:design}. The enclave is assigned to a dedicated logical core, while one or more logical cores are assigned to the untrusted part of the application. The code integrity inside an enclave can be authenticated through the SGX remote attestation feature mentioned in section~\ref{sgx-back}.

Initially, packets are received by the untrusted part of the application through one or more receiving (Rx) queues and are placed in DPDK buffers. The memory address (pointer) of each buffer is enqueued to a receiving (Rx) DPDK ring. The enclave repeatedly dequeues pointers from the Rx ring, so that the buffers are processed by a secure module within the enclave. Once processing is done, the buffer pointers are enqueued to a transmission (Tx) DPDK ring. The untrusted application part dequeues them from the transmission ring and transmits the buffers through one or more transmission (Tx) queues. We use the user-space DPDK I/O mechanisms to read buffers from the NICs and enqueue/dequeue them to/from the Rx, and Tx rings, since such libraries have been specifically optimized for high traffic rates.

This design can be used for programmable routing and switching applications. Our experimental results (section~\ref{l2forwarding}) indicate that such trusted application models based on this design can achieve forwarding performance of $\sim 22$ million packets/frames per second for 64-byte frames.

\subsection {Parallel Packet Processing Approach}
\label{scaling}

\begin{figure}[h]
\centering
\includegraphics[scale=0.31]{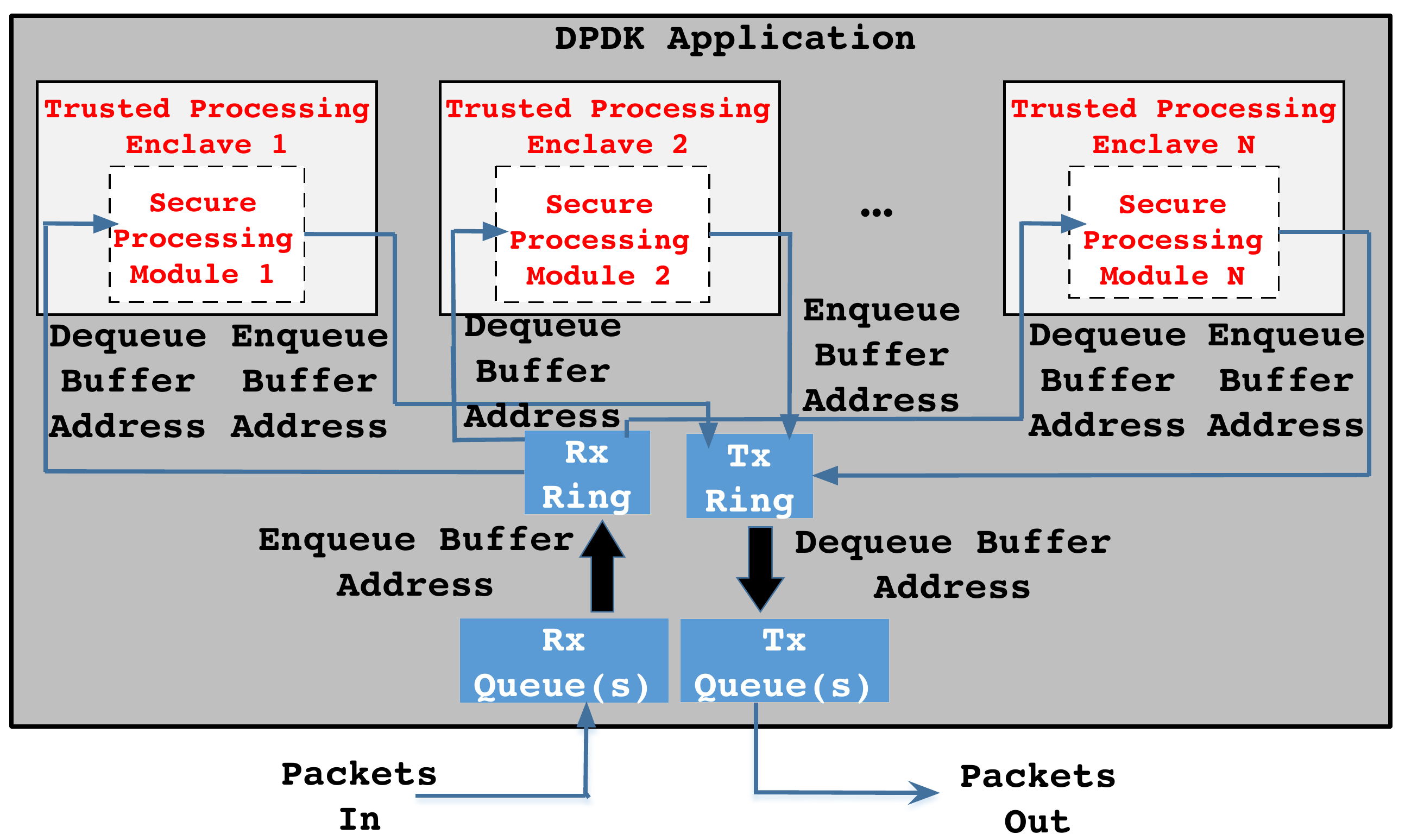}
\caption {Secure Parallel Packet Processing Design}
\label{Figure:design-scaling}
\end{figure}

Depending on the requirements of each application, an enclave might have to perform costly operations, which can result in considerable performance degradation. To enable applications to scale their performance in such cases, multiple processing enclaves that operate in parallel can be instantiated (transition from sequential to parallel processing). To this end, in Figure~\ref{Figure:design-scaling}, we present our design approach to scale packet processing by utilizing multiple SGX enclaves, which implement the same processing logic within the same DPDK application. The integrity of the code executed by each enclave can be authenticated through remote attestation. Similar to section~\ref{baseline}, a separate logical core is assigned to each enclave, while the untrusted part of the application may be assigned one or more logical cores. Each enclave dequeues buffer pointers from the Rx ring and processes the associated buffers. Finally, it enqueues the buffer pointers to the Tx ring for transmission.

This design can be used for multi-threaded applications and systems (e.g., load-balancers, spark, hadoop and other data processing systems) and, more general, in scenarios, where a single processing enclave results in low performance (e.g., because of EPC paging). Such example scenarios are the processing of encrypted L2 and L3 traffic (similar to MACSec and IPSec) presented in sections~\ref{encr-L2-forw} and~\ref{encr-L3-forw}, where this design achieves a performance gain of 3-4x compared to the baseline sequential design. 

%\begin{figure*}[t!]
%	\centersing
%	\begin{subfigure}[t]{0.5\textwidth}
%		\centering
%		\includegraphics[scale=0.315]{figures/design-scaling.pdf}
%		\caption {Secure Packet Processing Scaling Design}
%		\label{Figure:design-scaling}
%	\end{subfigure}%
%	~ 
%	\begin{subfigure}[t]{0.5\textwidth}
%		\centering
%		\includegraphics[scale=0.295]{figures/design-pipeline.pdf}
%		\caption {Secure Packet Processing Pipeline Design}
%		\label{Figure:design-pipeline}
%	\end{subfigure}
%	\caption{Secure Packet Processing Scaling \& Pipeline Design}
%\end{figure*}

\subsection {Packet Processing Pipeline Approach}
\label{pipeline}

\begin{figure}[h]
\centering
\includegraphics[scale=0.3]{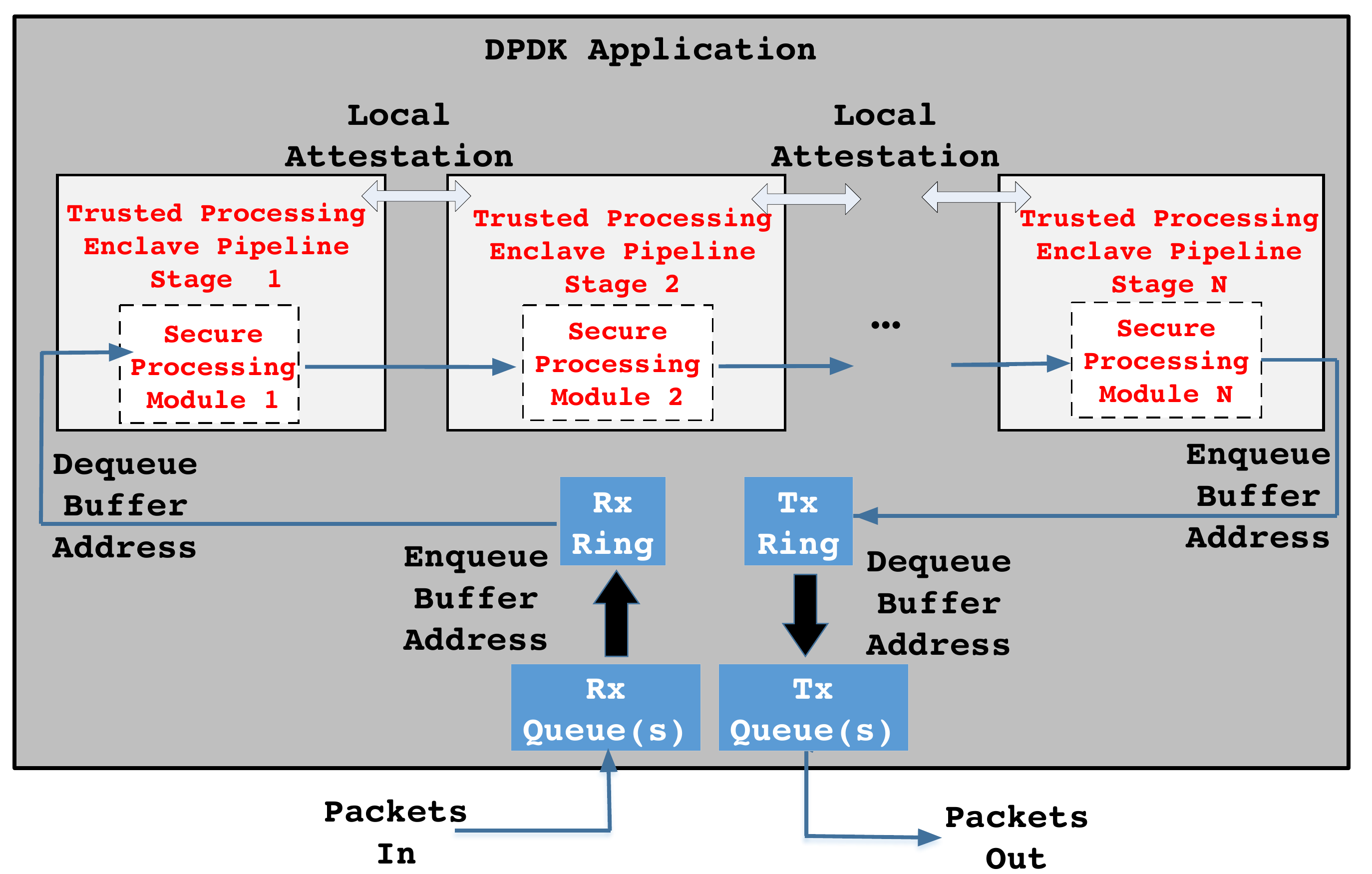}
\caption {Secure Packet Processing Pipeline Design}
\label{Figure:design-pipeline}
\end{figure}

In Figure~\ref{Figure:design-pipeline}, we present the design of a secure packet processing pipeline design consisting of a number of stages. Each stage applies a specific action to packets and forwards them to the subsequent processing stage, and is implemented as a separate SGX enclave. The code integrity of the first stage enclave is authenticated through remote attestation, while each next stage enclave and its previous one authenticates each other and establishes a protected communication channel through local attestation. Each stage enclave is assigned to a separate logical core.

The first stage enclave dequeues buffer pointers from the Rx ring and initiates secure processing. The result of the first stage processing will be the input of the second stage and so on, so forth. The final processing stage of the pipeline enqueues the processed buffer pointers to the Tx ring for transmission.

This design can be used in networking and systems environments that require an action sequence to be applied to received packets, such as Software Defined Networking (SDN) switches~\cite{morreale2015software}, Deep Packet Inspection (DPI) systems~\cite{abuhmed2008survey}, and cloud applications following the microservice architecture~\cite{nadareishvili2016microservice}. We expect that it would allow for more flexible and diverse processing than the previous approaches, but the performance evaluation of this design is left to our future work.

\subsection{Threat Model}

We focus on cases, where systems and applications may process confidential information, therefore they leverage trusted execution to secure crucial data and perform critical processing operations. Such cases can occur either in an untrusted cloud setup due to an accidental leak of confidential information or malicious tenants that try to compromise the execution integrity of others, or in setups, where resources are distributed at the edge of the network. Similar to the use-cases mentioned in section~\ref{threat-models}, attackers may attempt to learn routing and forwarding policies, compromise encryption and authentication keys, monitor encrypted traffic, etc. According to the SGX threat model~\cite{mckeen2013innovative}, we assume that an attacker can compromise software components, including privileged code (e.g., OS and BIOS) and launch physical attacks.

Following the SGX threat model~\cite{mckeen2013innovative} and prior related work~\cite{kim2017enhancing, kim2015first}, DoS and DDoS attacks are out of the scope of our work, since compromised software or hardware can deny the service at any point of the execution (e.g., restart or crash the system or flush all the unprotected DPDK memory). The same assumption applies to side channel attacks against SGX (e.g., page fault and cache-based side channel attacks). Software techniques to protect SGX-capable applications against attacks aiming to exploit bugs (e.g., buffer overflows, synchronization bugs, etc.)~\cite{weichbrodt2016asyncshock, seo2017sgx} are also out of the scope of our work.

\section {Implementation}
\label{implementation}

In this section, we describe implementation-specific details on our effort to combine DPDK and SGX, as well as the developed application scenarios.

\subsection{Combining DPDK and SGX}

We used DPDK version 17.02 and implemented our applications as enclaves using SGX version 1.8. To make the enclave ECALL/OCALL API accessible to DPDK context, we had to first compile the enclavized application and then compile the DPDK context, which was using the enclave API as a shared library. To be able to make use of more DPDK context in enclave mode, we had to modify the DPDK codebase to alleviate its coupling with standard C libraries (e.g., by modifying functions for log collection and printing, converting inline functions to macros), since SGX currently supports only the use of memory allocation and deallocation standard C libraries in enclave mode. However, in some cases, the coupling was so tight (e.g., DPDK libraries for hash and flow table implementation and lookup) that a specific OCALL had to be executed. We discuss further details and our optimization to alleviate this additional overhead for such cases in section~\ref{application-scenarios}.

The API of our application enclaves exposes a single ECALL, which is required to be called only once during the enclave instantiation. No further ECALLs are required during the execution, since the communication between DPDK and the enclave is achieved through DPDK rings, a structure optimized for performance. Most of the developed codebase, including all the performance-sensitive enclave components, is written in C instead of C++ for compatibility with the DPDK codebase and performance purposes.

\subsection{Application Scenarios}
\label {application-scenarios}

We implemented a few DPDK application scenarios to experiment with. The enclavized components of each application scenario are summarized in Table~\ref{Table:encl-comps}.

\textbf{Layer 2 (L2) forwarding}: A frame processing application implementing the operation of a network switch. We use a single processing enclave for the trusted applications following the design of section~\ref{baseline}. The untrusted application part receives frames, which enqueues to the Rx ring. The processing enclave dequeues frames from this ring and performs a number of frame sanity checks, destination MAC address lookup and source MAC address rewriting. Once processing is done, the enclave enqueues the frames to the Tx ring. 

\textbf{Layer 3 (L3) forwarding}: A packet processing application implementing the operation of a network router. The enclave operations include a few packet header sanity checks and the longest prefix match lookup of the destination address of a packet.

\textbf{Encrypted L2 forwarding}: A frame processing application for encrypted Ethernet traffic implementing the operation of a MACsec-capable switch. We use the frame format of MACsec~\cite{romanow2006media}, where the Integrity Check Value (ICV) field consists of an 128-byte CMAC hash, and multiple processing enclaves following the design of section~\ref{scaling}. The untrusted application part receives encrypted frames, which are enqueued to a processing enclave. A processing enclave dequeues and decrypts them, while it also generates the ICV of the raw frame data and compares it with the ICV of the received frames to verify their integrity. If the integrity verification is successful, the raw frames are processed (MAC table lookup and MAC address rewriting) and their new ICV is generated. Finally, the enclave encrypts the processed frames and enqueues them to the Tx ring to be forwarded.

\textbf{Encrypted L3 forwarding}: A packet processing application for encrypted network level traffic implementing the operation of a VPN endpoint (router). We use multiple processing enclaves following the design of section~\ref{scaling} and the packet format provided by the ''Encapsulating Security Payload'' function of IPsec~\cite{kent2005ip}, where the Integrity Check Value (ICV) field consists of an 128-byte CMAC hash. The workflow is similar to the one explained for the encrypted L2 forwarding application.

\textbf{Load-balancing \& backend server processing}: An application implementing the operation of a load balancer distributing traffic to multiple backend server processes (either VMs or containers) running on the same physical machine. The load-balancing process that maintains a flow table (with 1 million flow entries) and classifies the received packets into flows based on their destination IP address. The backend server processes filter and forward the distributed packets based on their destination IP address (hash-based forwarding). The load balancer forwards traffic to the backend processes through a number of DPDK rings (one ring per backend process).

An enclave of a load balancing or server process has to make an explicit OCALL for every batch of packets dequeued from the Rx ring. The vanilla DPDK application uses DPDK libraries for the flow table and the hash table lookups that leverage standard C libraries, which are not trusted and cannot be used in enclave mode. To allow the DPDK libraries to access the enclave buffer containing the keys for the lookups, this buffer has to be copied during the OCALL transition from the enclave's EPC to untrusted memory. To return the lookup results back to the enclave, a separate buffer has to be copied from the untrusted memory to the enclave's EPC when the OCALL returns. In addition to the copy operation itself, additional checks have to be performed by SGX to ensure that the full memory range of the buffers passed to the untrusted application part is within the enclave. 

To enhance system performance and alleviate overheads, we performed the following optimizations:

\begin{itemize}[leftmargin=*]

\item Increased the number of packets dequeued as batch by a server or load balancing enclave from its corresponding  ring to reduce the total number of performed OCALLs. 

\item Used untrusted memory for the buffers of the lookup keys and results to avoid memory copy from/to the EPC and the additional checks. Essentially, we traded security for performance, as explained in section~\ref{lessons}.

\item According to the design presented in section~\ref{scaling}, we enabled each backend process to instantiate two enclaves implementing the same processing logic.

\end{itemize}

\begin{figure}[h]
\centering
\includegraphics[scale=0.5]{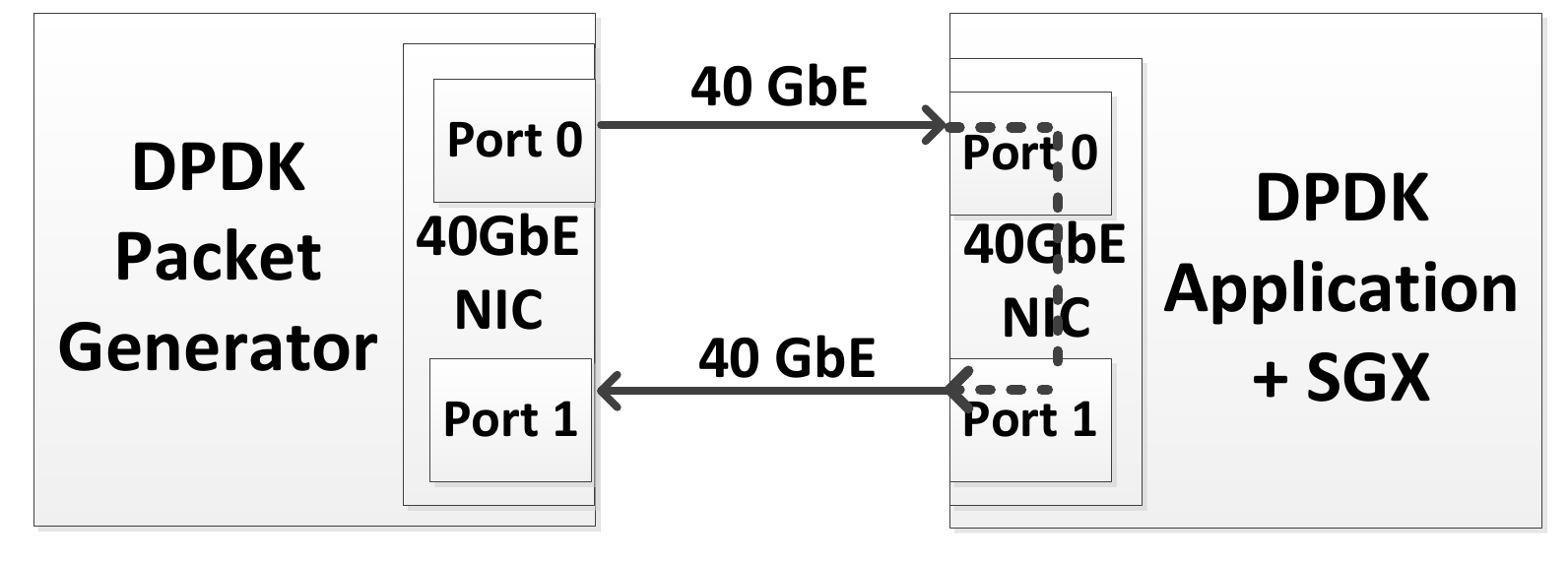}
\caption {Experimental Setup}
\label{Figure:setup}
\end{figure}

\begin{table*}[h!]
	\centering
	\caption{Enclavized Components For Each Application Scenario}
	\label{Table:encl-comps}
	\begin{tabular}{|c|c|}
		\hline
		\textbf{Application}              & \textbf{Enclavized Components}                                                                                                                                                              \\ \hline
		L2 Forwarding (Plain Traffic)     & \begin{tabular}[c]{@{}c@{}}Sanity check, MAC address lookup, \\ MAC address rewriting\end{tabular}                                                                                          \\ \hline
		L3 Forwarding (Plain Traffic)     & \begin{tabular}[c]{@{}c@{}}Sanity Check, longest prefix match lookup \end{tabular}                                                                                  \\ \hline
		L2 Forwarding (Encrypted Traffic) & \begin{tabular}[c]{@{}c@{}}Frame decryption, sanity check, frame integrity check, \\ MAC address lookup, MAC address rewriting, \\ ICV generation, frame encryption\end{tabular}            \\ \hline
		L3 Forwarding (Encrypted Traffic) & \begin{tabular}[c]{@{}c@{}}Packet decryption, sanity check, packet integrity check, \\ longest prefix match lookup, ICV generation, packet encryption\end{tabular} \\ \hline
		Load Balancer                     & \begin{tabular}[c]{@{}c@{}}Packet processing before and after \\ flow table lookup\end{tabular}                                                                                             \\ \hline
		Back-end Server                   & \begin{tabular}[c]{@{}c@{}}Packet processing before and after \\ hash table lookup for hash-based forwarding\end{tabular}                                                                   \\ \hline
	\end{tabular}
\end{table*}

\section {Experimental Evaluation}
\label{evaluation}

We first discuss our experimental setup, methodology and workload and then present the results of our study to evaluate the performance and the SGX related overhead of our design for the application scenarios mentioned in Section~\ref{application-scenarios}.

\subsection {Experimental Setup}

\textbf{Testbed:} For our experimental evaluation, we use a testbed consisting of 2 machines (Figure~\ref{Figure:setup}). On the first one, we run the DPDK packet generator (version 3.2) and on the second one, the application under test (trusted and vanilla DPDK applications). Each machine is equipped with an Intel Xeon CPU E3-1240L v5 (2.10GHz, 4 cores), 32GB of RAM, and a dual-port Intel XL710 40GbE NIC. The NIC is connected to the processor on each machine through a PCIe Gen3 x8 bus. Each of port 0 and port 1 of the generator is connected to port 0 and port 1 respectively of the application under test through a 40 GbE Ethernet link. 

\textbf{Methodology:} The packet generator creates traffic of variable size (64 to 1500-byte frames or up to the size that saturates the available link capacity) and sends it through port 0 of its NIC to port 0 of the application's NIC. The application processes the received traffic and forwards it through port 1 of its NIC to port 1 of the packet generator's NIC. The performance is quantified by measuring the received packet rate in Million Packets Per Second (MPPS) and throughput in Gbps at port 1 of the packet generator's NIC. We run each experiment 10 times (each experiment's duration is 1 minute) and we report on the average results among the runs, since the run-to-run result variation is negligible.

To calculate the SGX overhead, we use the following equation, where the ``Vanilla App (MPPS)'' term refers to the measured MPPS processed by the vanilla DPDK application and the ``SGX App (MPPS)'' term to the measured MPPS processed by the trusted DPDK application:

\begin{equation*}
\begin{split}
Overhead &(\%) = \\
&\frac{Vanilla App (MPPS) - SGX App (MPPS)}{Vanilla App (MPPS)} * 100
\end{split}
\end{equation*}

\textbf{Workload:} For this initial evaluation of our design, we use a synthetic traffic workload rather than real-world traffic traces for 2 reasons: 1) to study the impact of gradually greater sizes of traffic patterns to the SGX performance, and 2) to stress on the specific fields and layer of the TCP/IP protocol stack that processing and forwarding are based on for each scenario. For the L3 related and the load balancing \& server processing scenarios, our workload consists of traffic flows with 1 million distinct source and destination IP addresses, and the MAC address of the generator's and application's port 0 as source and destination respectively. For the L2 related scenarios, our workload consists of 1 million distinct source and destination MAC addresses.

%The reason that we did not use real-world traffic traces, but rather a synthentic workload, is that for our initial evaluation, we aimed to stress on the specific fields and layer of the TCP/IP protocol stack that processing and forwarding was based on for each application scenario to examine the SGX related overhead under diverse processing conditions.

\subsection {Experimental Results}
\label{results}

%In this section, we present our experimental results based on the following DPDK applications: 1) Layer 2 (L2) forwarding (frame processing), 2) Layer 3 (L3) forwarding (packet processing), 3) Encrypted L2 forwarding, 4) Encrypted L3 forwarding, and 5) Load balancing \& server processing.

\subsubsection {L2 \& L3 Forwarding}
\label{l2forwarding}

\begin{figure}[h]
\centering
\includegraphics[width=\columnwidth]{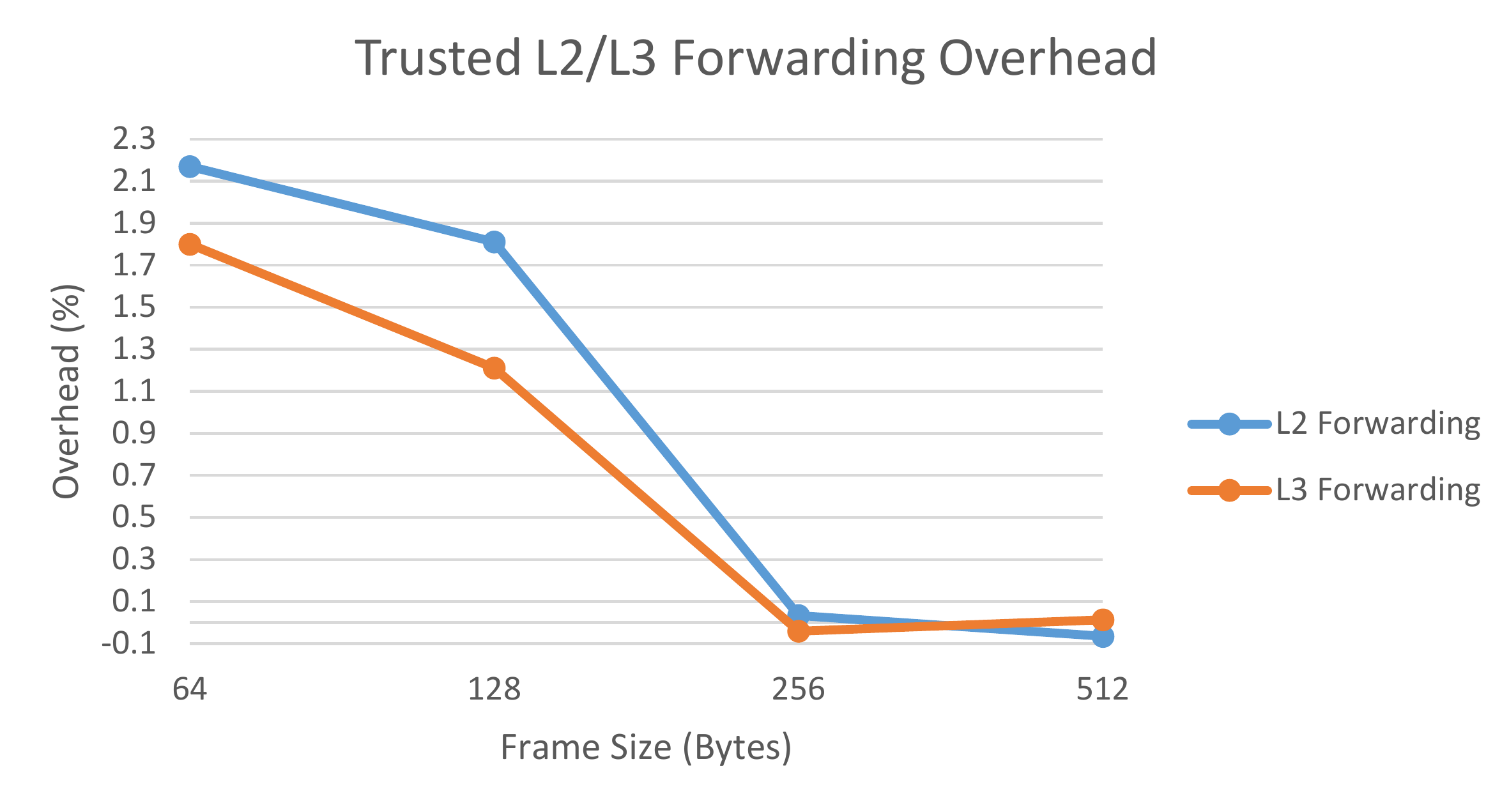}
\caption {Trusted L2 \& L3 Forwarding Overhead}
\label{Figure:L2-L3-over}
\end{figure}

In Figure~\ref{Figure:L2-L3-over}, we present the performance overhead of a trusted L2 and L3 forwarding application compared to the corresponding vanilla DPDK L2 and L3 forwarding application. For 64-byte frames, the overhead is $\sim 2.1\%$ and $\sim 1.8\%$, while it decreases to $\sim 1.8\%$ and $\sim 1.2\%$ for 128-byte frames respectively. When we increase the frame size to 256 and 512 bytes, the transmission and reception delays become the bottleneck of the overall system, therefore, the performance of the trusted application reaches the performance of the vanilla one.

In Table~\ref{Table:Throughput}, we present the performance (in MPPS) and the achieved wire throughput (determined by the illustrated frame size plus 8 bytes of preamble and 12 bytes of interframe gap) for both the vanilla and the trusted applications. For small-sized frames, we cannot saturate the available link capacity, because the NICs are limited by the current CPU power per core, which is not sufficient for our network bandwidth. Similar results have been presented by DPDK performance evaluation studies with similar setup~\cite{kawashima2016host, kawashima2017evaluation}. As the frame size increases, our applications are required to forward fewer frames, while we approach the saturation of the available bandwidth for 512-byte frames.

%\begin{figure}[h]
%\centering
%\includegraphics[width=\columnwidth]{figures/L2-over.pdf}
%\caption {Trusted L2 Forwarding Overhead}
%\label{Figure:L2-over}
%\end{figure}

%\begin{figure}[h]
%\centering
%\includegraphics[width=\columnwidth]{figures/L2-perf.pdf}
%\caption {L2 Forwarding Performance}
%\label{Figure:L2-perf}
%\end{figure}

%\begin{figure}[h]
%\centering
%\includegraphics[width=\columnwidth]{figures/L2-band.pdf}
%\caption {L2 Forwarding Wire Throughput}
%\label{Figure:L2-band}
%\end{figure}

\begin{table*}[h!]
\centering
\caption{L2 \& L3 Forwarding Performance \& Wire Throughput}
\label{Table:Throughput}
\begin{tabular}{|c|c|c|c|c|}
\hline
\textbf{Frame Size (Bytes)}             & \multicolumn{4}{c|}{\textbf{Performance (MPPS)/Wire Throughput (Gbps)}}                                                                           \\ \hline
\multicolumn{1}{|l|}{\multirow{2}{*}{}} & \multicolumn{2}{c|}{L2 Forwarding}                          & \multicolumn{2}{c|}{L3 Forwarding}                          \\ \cline{2-5} 
\multicolumn{1}{|l|}{}                  & \multicolumn{1}{c|}{Vanilla} & \multicolumn{1}{c|}{Trusted} & \multicolumn{1}{c|}{Vanilla} & \multicolumn{1}{c|}{Trusted} \\ \hline
64                                      & 21.80/14.65                        & 21.33/14.33                        & 21.89/14.71                        & 21.50/14.45                        \\ \hline
128                                     & 21.34/25.30                         & 20.95/24.80                        & 21.45/25.30                         & 21.19/25.09                        \\ \hline
256                                     & 13.75/30.37                        & 13.75/30.36                        & 13.83/30.54                        & 13.84/30.56                        \\ \hline
512                                     & 8.67/36.91                        & 8.68/36.93                        & 8.68/36.93                        & 8.68/36.93                        \\ \hline
\end{tabular}
\end{table*}

\subsubsection {Encrypted L2 Forwarding}
\label{encr-L2-forw}

In this scenario, we vary the number of enclaves used for the processing of encrypted L2 traffic. Our goal through this experiment is not to replicate the cipher suite used by MACsec or the specifics of the protocol, but rather study a trusted application that uses the mechanisms provided by SGX for encryption/decryprtion.

\begin{figure}[h]
\centering
\includegraphics[width=\columnwidth]{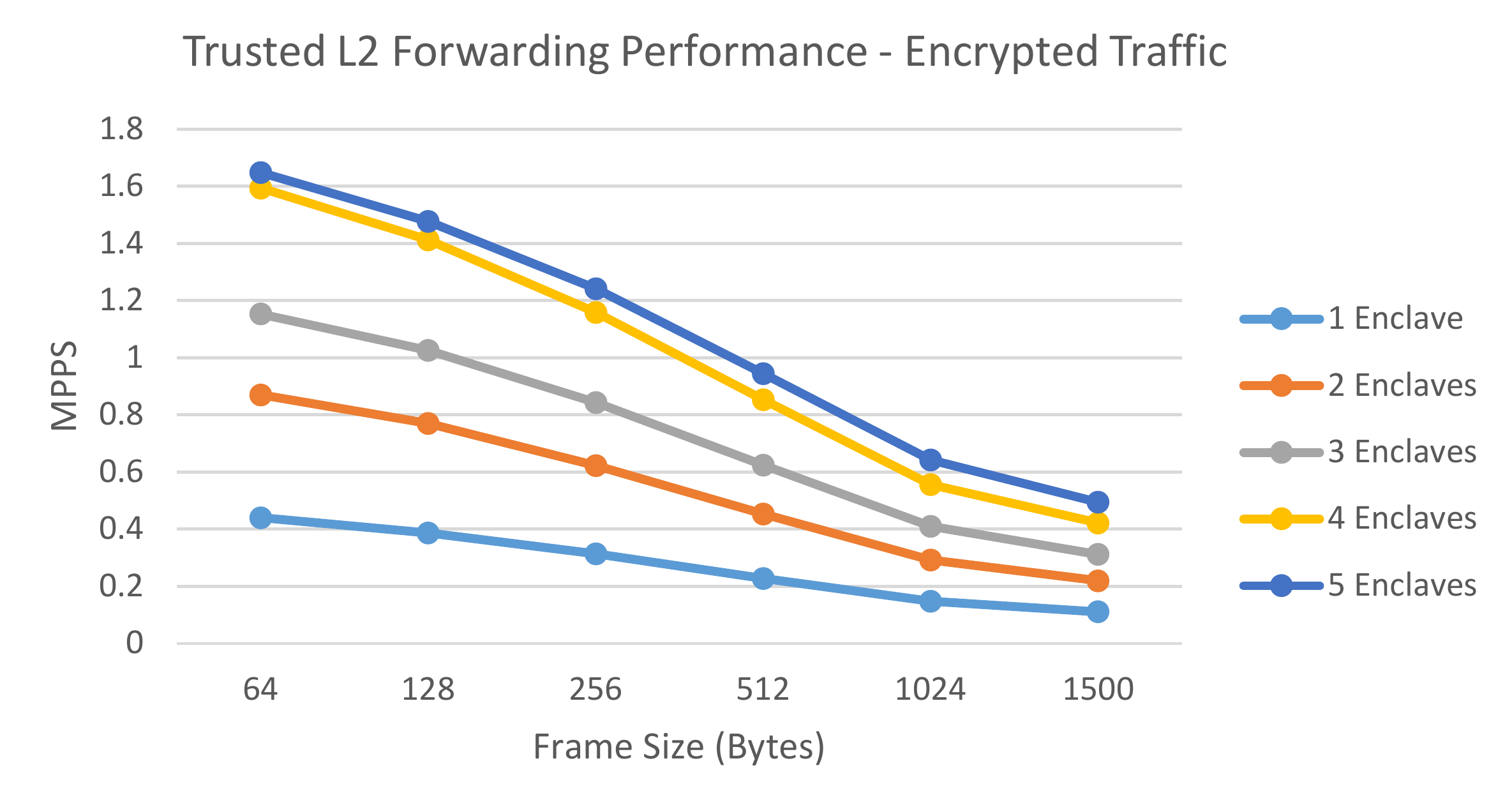}
\caption {Trusted L2 Forwarding Performance -- Encrypted Traffic}
\label{Figure:L2-enc-perf}
\end{figure}

\begin{figure}[h]
\centering
\includegraphics[width=\columnwidth]{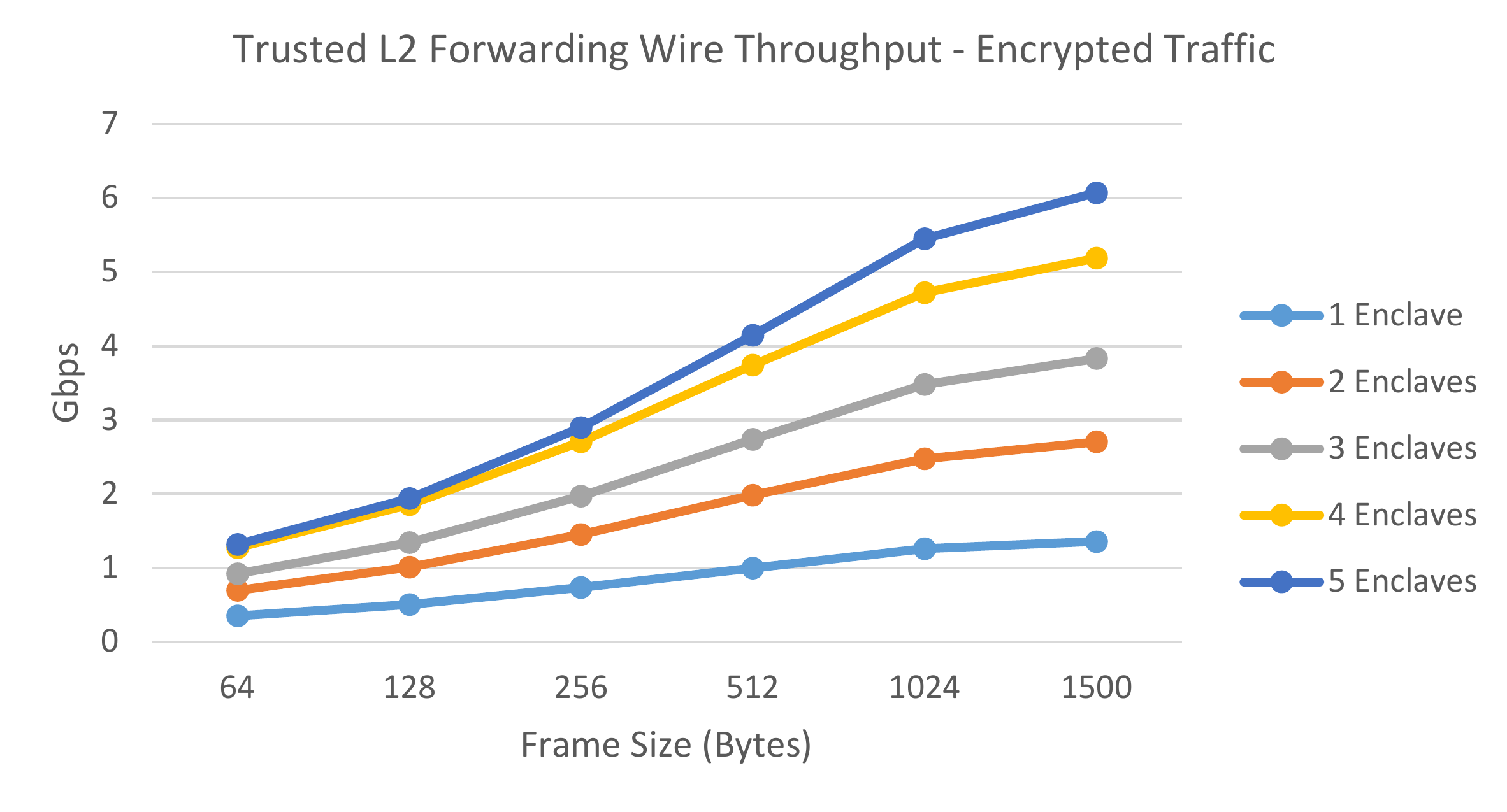}
\caption {Trusted L2 Forwarding Wire Throughput -- Encrypted Traffic}
\label{Figure:L2-enc-band}
\end{figure}

\begin{figure}[h]
\centering
\includegraphics[width=\columnwidth]{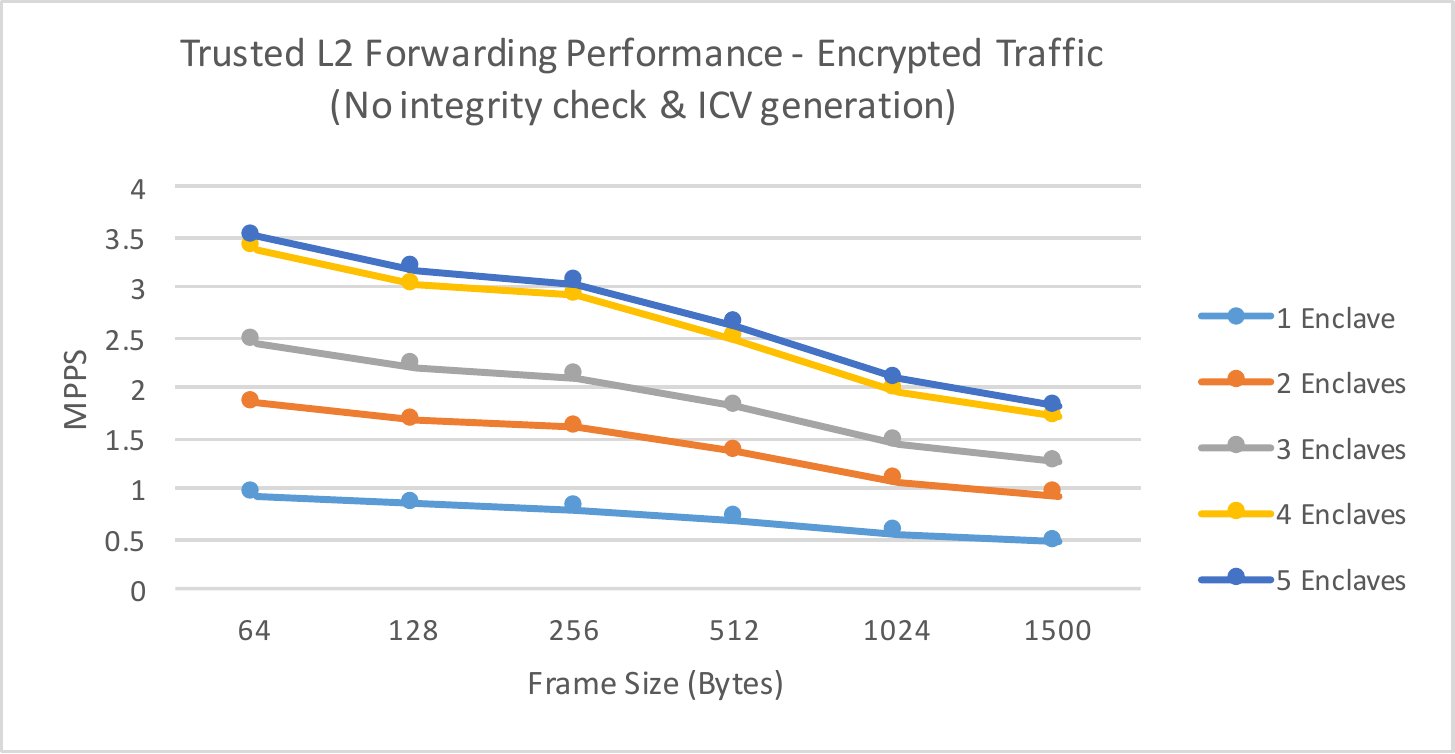}
\caption {Trusted L2 Forwarding Performance -- Encrypted Traffic (No Integrity Verification \& ICV Generation)}
\label{Figure:L2-enc-perf-no-int-ver}
\end{figure}

\begin{figure}[h]
\centering
\includegraphics[width=\columnwidth]{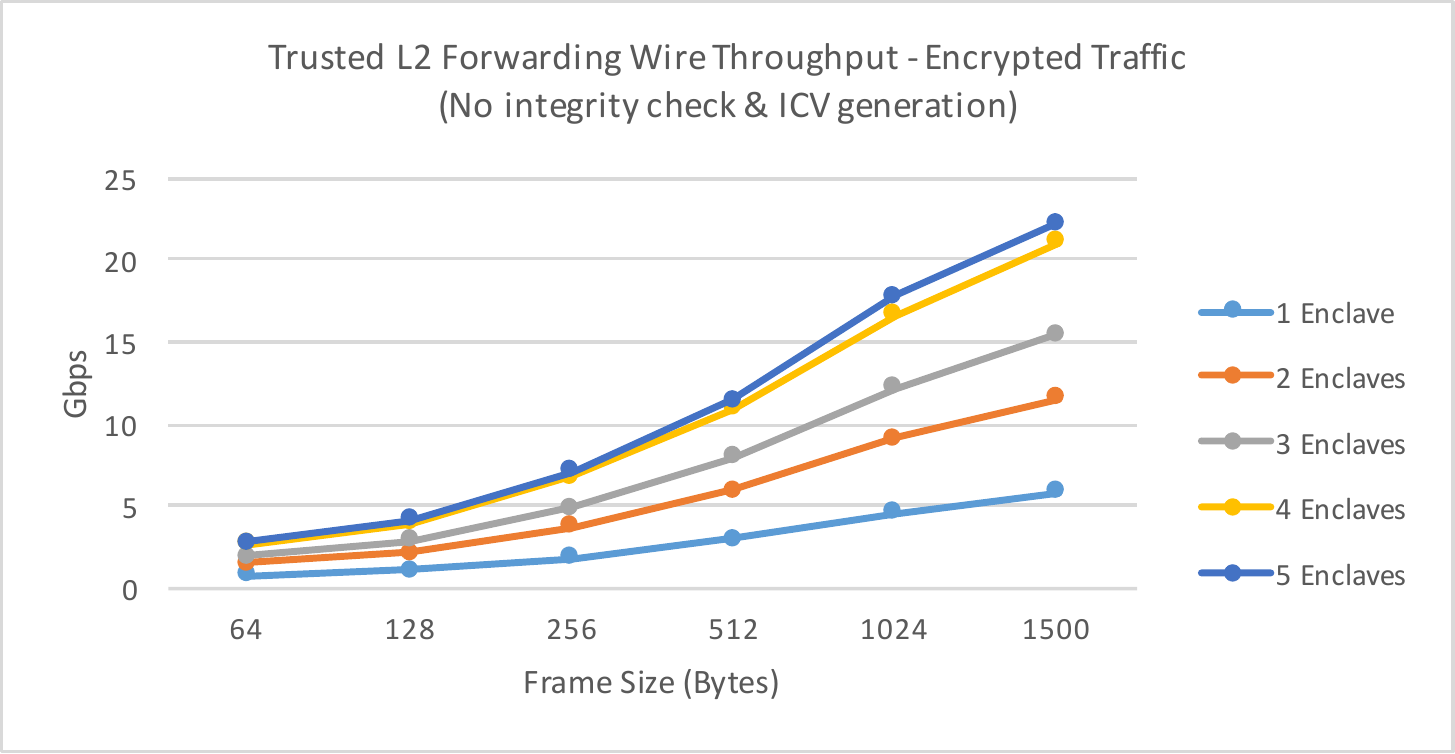}
\caption {Trusted L2 Forwarding Wire Throughput  -- Encrypted Traffic (No Integrity Verification \& ICV Generation)}
\label{Figure:L2-enc-band-no-int-ver}
\end{figure}

In Figure~\ref{Figure:L2-enc-perf} and~\ref{Figure:L2-enc-band}, we present the performance (MPPS) and the wire throughput (Gbps) respectively of trusted L2 forwarding for encrypted traffic. The results show that the system performance is low, however, it scales as we increase the number of enclaves. The potential bottlenecks in this scenario can be: 1) the SGX encryption/decryption overhead, and 2) the integrity verification process and the new ICV generation. To further investigate these bottlenecks, we repeated the same experiment without performing integrity verification and ICV generation to focus on the overhead of the SGX encryption/decryption. The results are presented in Figure~\ref{Figure:L2-enc-perf-no-int-ver} and~\ref{Figure:L2-enc-band-no-int-ver}. We observe that we achieve a speedup of $\sim 2.2x$, however, the results do not approach the ones for plain traffic. Therefore, we can conclude that the performance is dominated by the SGX encryption/decryption overhead.

\subsubsection {Encrypted L3 Forwarding}
\label{encr-L3-forw}

\begin{figure}[h]
\centering
\includegraphics[width=\columnwidth]{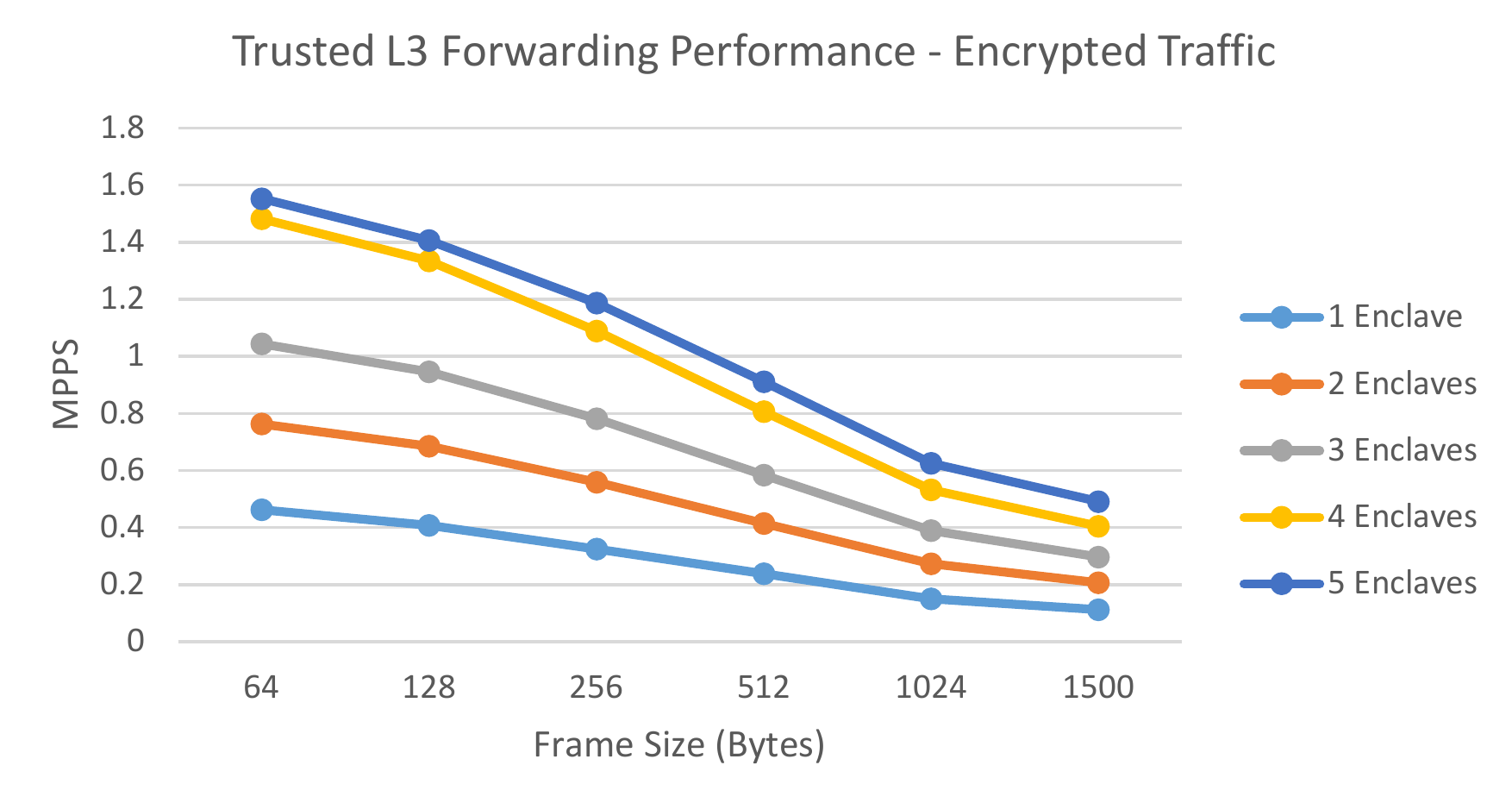}
\caption {Trusted L3 Forwarding Performance -- Encrypted Traffic}
\label{Figure:L3-enc-perf}
\end{figure}

\begin{figure}[h]
\centering
\includegraphics[width=\columnwidth]{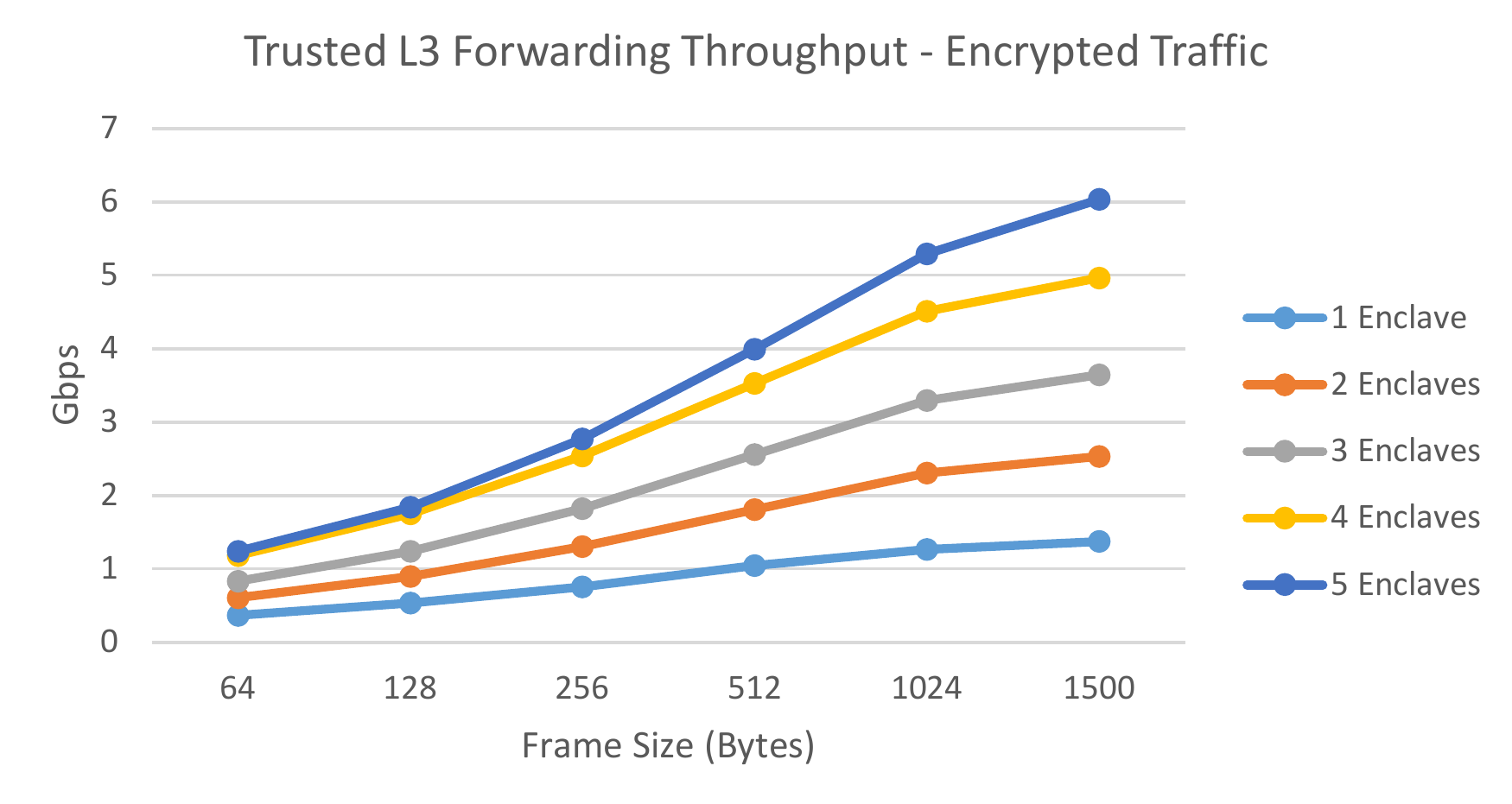}
\caption {Trusted L3 Forwarding Wire Throughput -- Encrypted Traffic}
\label{Figure:L3-enc-band}
\end{figure}

In this scenario, we vary the number of enclaves used for the processing of encrypted L3 traffic. In Figure~\ref{Figure:L3-enc-perf} and~\ref{Figure:L3-enc-band}, we present the performance (MPPS) and the achieved wire throughput (Gbps) respectively. The performance is again low, but it scales as we increase the number of enclaves. To further investigate the performance bottleneck, we repeated the same experiment without performing integrity verification and ICV generation. The results were similar to section~\ref{encr-L2-forw}, achieving the same speedup of $\sim 2.2x$, and leading us to the conclusion that the system bottleneck is again the SGX encryption/decryption.

\subsubsection {Load Balancing \& Backend Server Processing}
\label{lb}

%represents a simplified datacenter networking scenario and

\begin{figure}[h]
\centering
\includegraphics[width=\columnwidth]{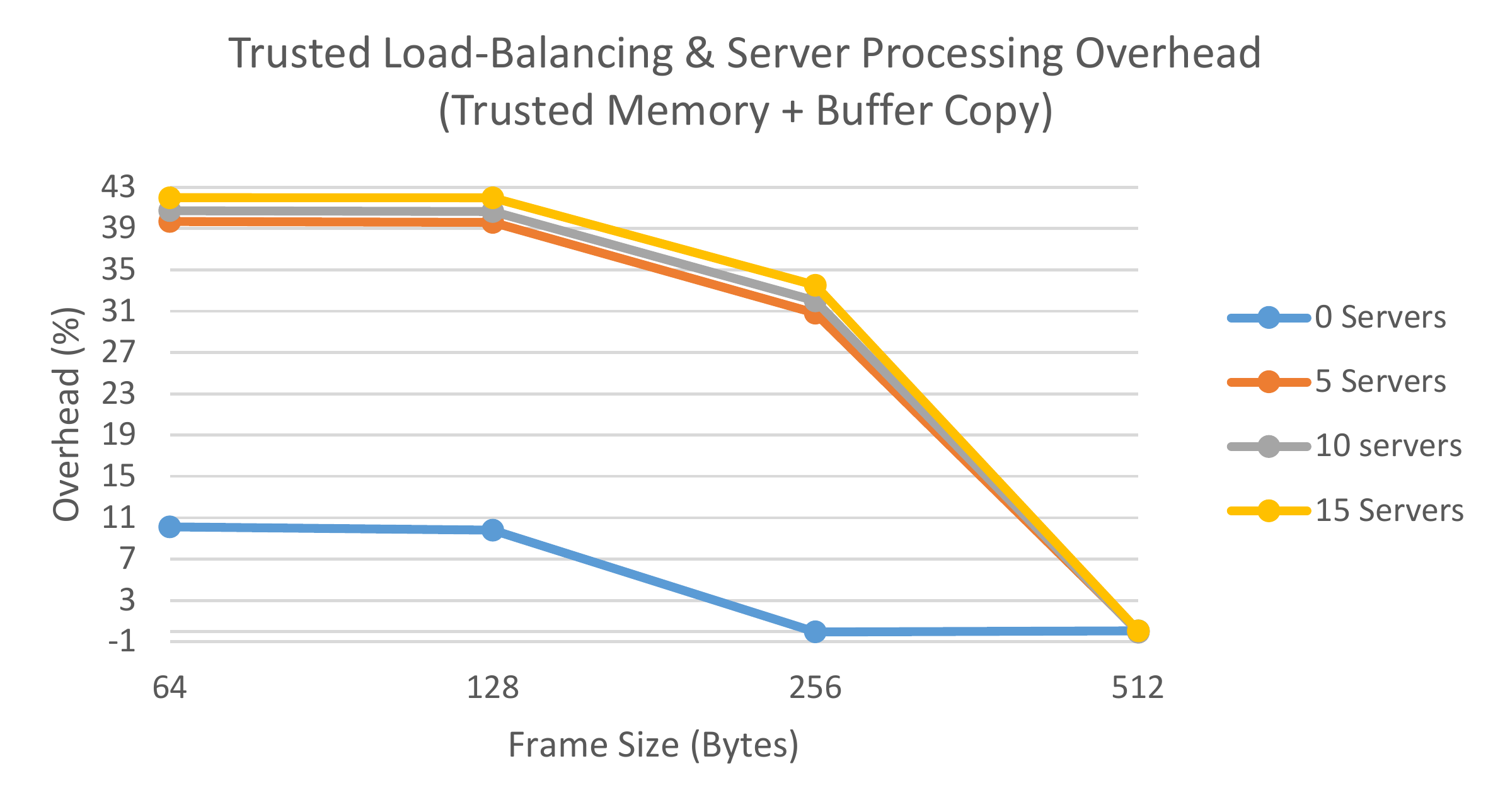}
\caption {Trusted Load Balancing \& Server Processing Overhead (Trusted Memory + Buffer Copy)}
\label{Figure:ld-over-copy}
\end{figure}

\begin{figure}[h]
\centering
\includegraphics[width=\columnwidth]{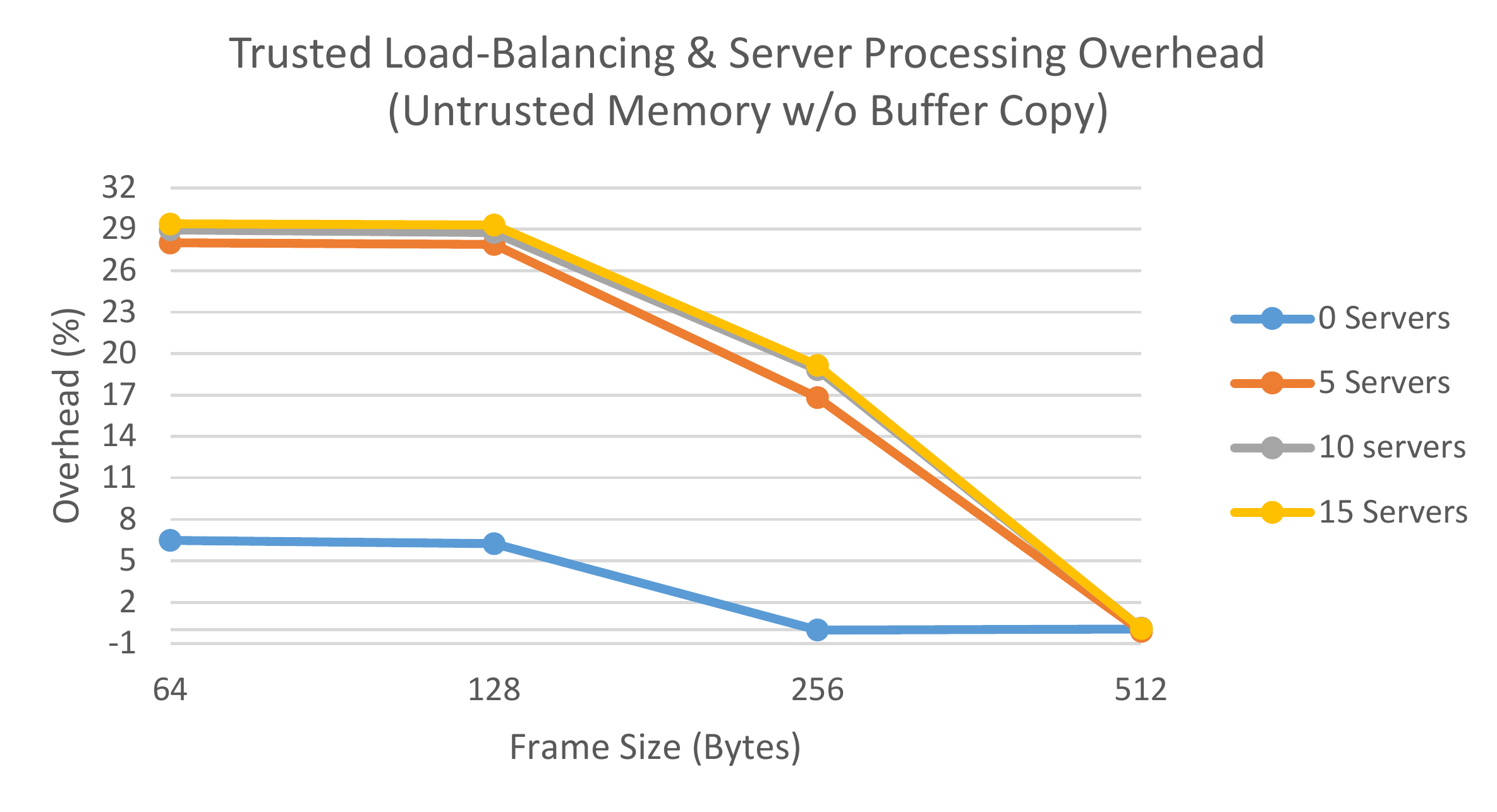}
\caption {Trusted Load Balancing \& Server Processing Overhead (Untrusted Memory w/o Buffer Copy)}
\label{Figure:ld-over-no-copy}
\end{figure}

In Figure~\ref{Figure:ld-over-copy} and~\ref{Figure:ld-over-no-copy}, we present the overhead of using trusted and untrusted memory for the lookup key and result buffers respectively (i.e., with and without copying the buffer memory during OCALLs). For trusted memory, 4 bytes are copied from EPC to untrusted memory and 1 byte from untrusted memory to EPC per packet for load balancing and 4 bytes from EPC to untrusted memory and vice versa per packet for server processing.

We observe that for load balancing only (no server processes running), the SGX overhead is $\sim 10.1\%$ and $\sim 6.5\%$ for 64-byte frames, and slightly decreases to $\sim 9.7\%$ and $\sim 6.2\%$ for 128-byte frames  (use of trusted and untrusted memory respectively). As we increase the number of server processes, the overhead increases and the performance degrades, since more enclaves have to run on the same physical machine and the number of performed OCALLs per dequeued packet batch doubles (one OCALL by the load balancing enclave and one by the corresponding server enclave). 

%In general, the overhead is lower when untrusted memory is used.

%The main reason for the overhead in this application scenario is that the enclave of a load balancing or server process has to make an explicit OCALL for every batch of packets dequeued for processing. More specifically, the vanilla DPDK application uses DPDK libraries for the flow table and the hash table implementations that leverage standard C libraries, which are not trusted and cannot be used in enclave mode.

%To optimize the overall system performance, we had to increase the number of packets dequeued as batch by a server Enclave from its corresponding Rx ring, so that we perform fewer OCALLs in total, and allow each server process to instatiate two enclaves performing the same processing instead of a single one (similar to the design presented in Section~\ref{scaling}).  

In Tables~\ref{Table:lb1} and~\ref{Table:lb2}, we present the performance (MPPS) and the achieved wire throughput (Gbps) for both the vanilla and trusted applications. As we increase the frame size, our applications are able to process fewer frames, while the achieved throughput approaches the maximum link capacity.

% \usepackage{multirow}
\begin{table*}[h!]
\centering
\caption{Load Balancing \& Backend Server Processing - Performance \& Wire Throughput (0 \& 5 Servers)}
\label{Table:lb1}
\begin{tabular}{|c|c|c|c|c|c|c|}
\hline
\textbf{Frame Size (Bytes)}             & \multicolumn{6}{c|}{\textbf{Performance (MPPS) / Wire Throughput (Gbps)}}                                                                                                                                                                                                                                               \\ \hline
\multicolumn{1}{|l|}{\multirow{2}{*}{}} & \multicolumn{3}{c|}{0 Servers}                                                                                                                             & \multicolumn{3}{c|}{5 Servers}                                                                                                                             \\ \cline{2-7} 
\multicolumn{1}{|l|}{}                  & Vanilla     & \begin{tabular}[c]{@{}c@{}}Trusted \\ (No Buffer \\ Copy)\end{tabular} & \begin{tabular}[c]{@{}c@{}}Trusted \\ (Buffer \\ Copy)\end{tabular} & Vanilla     & \begin{tabular}[c]{@{}c@{}}Trusted \\ (No Buffer \\ Copy)\end{tabular} & \begin{tabular}[c]{@{}c@{}}Trusted \\ (Buffer \\ Copy)\end{tabular} \\ \hline
64                                      & 17.51/11.76 & 16.37/11.00                                                            & 15.73/10.57                                                         & 16.51/11.09 & 11.88/7.98                                                             & 9.96/6.69                                                           \\ \hline
128                                     & 17.38/20.58 & 16.31/19.31                                                            & 15.68/18.56                                                         & 16.39/19.41 & 11.83/14.01                                                            & 9.90/11.72                                                          \\ \hline
256                                     & 14.17/31.29 & 14.18/31.30                                                            & 14.18/33.31                                                         & 14.17/31.30 & 11.79/26.03                                                            & 9.80/21.65                                                          \\ \hline
512                                     & 8.99/38.26  & 8.98/38.23                                                             & 8.98/38.24                                                          & 8.99/38.28  & 8.99/38.27                                                             & 8.99/38.27                                                          \\ \hline
\end{tabular}
\end{table*}

% \usepackage{multirow}
\begin{table*}[h!]
\centering
\caption{Load Balancing \& Backend Server Processing - Performance \& Wire Throughput (10 \& 15 Servers)}
\label{Table:lb2}
\begin{tabular}{|c|c|c|c|c|c|c|}
\hline
\textbf{Frame Size (Bytes)}             & \multicolumn{6}{c|}{\textbf{Performance (MPPS) / Wire Throughput (Gbps)}}                                                                                                                                                                                                                                               \\ \hline
\multicolumn{1}{|l|}{\multirow{2}{*}{}} & \multicolumn{3}{c|}{10 Servers}                                                                                                                             & \multicolumn{3}{c|}{15 Servers}                                                                                                                             \\ \cline{2-7} 
\multicolumn{1}{|l|}{}                  & Vanilla     & \begin{tabular}[c]{@{}c@{}}Trusted \\ (No Buffer \\ Copy)\end{tabular} & \begin{tabular}[c]{@{}c@{}}Trusted \\ (Buffer \\ Copy)\end{tabular} & Vanilla     & \begin{tabular}[c]{@{}c@{}}Trusted \\ (No Buffer \\ Copy)\end{tabular} & \begin{tabular}[c]{@{}c@{}}Trusted \\ (Buffer \\ Copy)\end{tabular} \\ \hline
64                                      & 16.43/11.04 & 11.68/7.85                                                            & 9.74/6.55                                                         & 16.37/11.00 & 11.56/7.77                                                             & 9.50/6.38                                                           \\ \hline
128                                     & 16.32/19.33 & 11.64/13.79                                                            & 9.69/11.47                                                         & 16.30/19.30 & 11.52/13.64                                                            & 9.46/11.20                                                          \\ \hline
256                                     & 14.17/31.30 & 11.50/25.40                                                            & 9.64/21.29                                                         & 14.17/31.29 & 11.46/25.30                                                            & 9.42/20.81                                                          \\ \hline
512                                     & 8.99/38.30  & 8.99/38.30                                                             & 8.99/38.30                                                          & 8.99/38.25  & 8.98/38.22                                                             & 8.98/38.23                                                          \\ \hline
\end{tabular}
\end{table*}

\section {Discussion}
\label {discussion}

In this section, we summarize the lessons we have learned through our work, and identify a number of open issues that we are planning to address in our future work.

\subsection {Lessons Learned}
\label {lessons}

Our work highlighted a number of important technical points that we would like to share with the broader community. There is a clear trade-off between building a secure and trusted system and the performance that this system can achieve even in cases that execution integrity is directly integrated into its architecture. There are many different threat models and attack scenarios; building a system that provides confidentiality and security against multiple attack scenarios at the same time comes with a performance penalty. 

%The extent of this penalty depends on the requirements and nature of each system, as well as the applications and services running on it.

Our experimental evaluation indicted that there is a considerable ECALL/OCALL transition performance overhead. Transferring code execution from/to enclave mode to/from the untrusted application part does not come for free, since multiple hardware-level checks have to be performed. For example, as described in section~\ref{lb}, enclavized applications that require access to standard C libraries (other than the memory allocation and deallocation libraries, which are the only ones supported by SGX in enclave mode) have to invoke explicit OCALLs, which impacts their performance. To achieve better amortized transition performance, the number of ECALLs and OCALLs should be minimized and ensure that the application executes work inside the enclave for as long as possible. 

Multiple checks are also performed in the case of memory copy to/from EPC during ECALLs and OCALLs. As shown in section~\ref{lb}, such checks impose an additional performance penalty of 5-10\%. One way to overcome this penalty is the use of untrusted memory by an enclave to avoid copying memory from/to EPC for ECALLs and OCALLs which, however, raises certain security concerns; the data in the untrusted memory is not encrypted and can potentially be accessible by malicious actors, thus leaking enclave secrets, while it can also be altered at any point of the execution without notice.

\subsection {Open Issues}
\label{open-issues}

In our current work, we focused on evaluating the performance and trade-offs of the baseline and packet processing scaling design approaches (sections~\ref{baseline} and~\ref{scaling}). The evaluation of our packet processing pipeline approach (section~\ref{pipeline}), which could be used for the design of SDN-capable switches, DPI systems and cloud applications using the microservice architecture, is left to our future work. To conduct further evaluation, we would also like to use workloads consisting of real-world traffic traces, such as the CAIDA anonymized Internet traces~\cite{caida} for backbone Internet traffic, and the IMC 2010 Data Center Measurements~\cite{benson2010network} for cloud and data center traffic, more trusted server applications, such as Hadoop and Spark, and traffic loading balancing algorithms that could boost performance~\cite{randles2010comparative, cardellini1999dynamic}. 

To eliminate the need of invoking explicit OCALLs to access standard C libraries, we are planning to implement trusted standard C-like libraries, which should boost the overall system performance. In cases that the enclave size and the required protected data and software exceeds the maximum EPC size, exit-less services~\cite{orenbach2017eleos} can be used to alleviate the enclave exit overhead imposed by EPC paging. This is achieved by creating a secure virtual memory abstraction that implements application-level paging inside the enclave.
\section {Conclusion}
\label{conclusions}

In this paper, we presented proof-of-concept designs of TEEs for NFs and server applications deployed on the untrusted cloud. Our designs are based on Intel SGX, which we combine with DPDK to prototype high-performance trusted applications. Through this work, we learned several valuable lessons and identified remaining open issues, which we shared with the broader community. Our experimental evaluation showed that NFs involving plain traffic can achieve close to native performance, while NFs involving encrypted traffic and server processing can still achieve competitive performance. 

%\section {Acknowledgements}

%We would like to thank Intel's SGX architect, Simon P. Johnson, for providing great feedback and insight into the SGX design and architecture, and Intel's research scientist, Michael Kaminsky, for the fruitful discussion.

%\bibliography{refs}
%\bibliographystyle{plain}
\bibliographystyle{plain}
\bibliography{refs}

\begin{thebibliography}{10}

\bibitem{leak2}
7 most infamous cloud security breaches.
\newblock
  https://www.storagecraft.com/blog/7-infamous-cloud-security-breaches/.

\bibitem{caida}
The caida anonymized internet traces 2012 dataset.
\newblock https://www.caida.org/data/passive/passive\_2012\_dataset.xml.

\bibitem{leak1}
Massive cloud leak exposes alteryx, experian, us census bureau data.
\newblock
  https://www.darkreading.com/cloud/massive-cloud-leak-exposes-alteryx-experian-us-census-bureau-data/d/d-id/1330673?

\bibitem{leak3}
Personal data of 6 million verizon customers was leaked.
\newblock
  http://fortune.com/2017/07/13/verizon-personal-data-customers-leaked/.

\bibitem{abuhmed2008survey}
Tamer AbuHmed, Abedelaziz Mohaisen, and DaeHun Nyang.
\newblock A survey on deep packet inspection for intrusion detection systems.
\newblock {\em arXiv preprint arXiv:0803.0037}, 2008.

\bibitem{arasu2014querying}
Arvind Arasu, Ken Eguro, Raghav Kaushik, and Ravishankar Ramamurthy.
\newblock Querying encrypted data.
\newblock In {\em Proceedings of the 2014 ACM SIGMOD international conference
  on Management of data}, pages 1259--1261. ACM, 2014.

\bibitem{arm2009security}
ARM ARM.
\newblock Security technology building a secure system using trustzone
  technology (white paper).
\newblock {\em ARM Limited}, 2009.

\bibitem{aublintalos}
Pierre-Louis Aublin, Florian Kelbert, Dan O'Keeffe, Divya Muthukumaran,
  Christian Priebe, Joshua Lind, Robert Krahn, Christof Fetzer, David Eyers,
  and Peter Pietzuch.
\newblock Talos: Secure and transparent tls termination inside sgx enclaves.

\bibitem{baumann2015shielding}
Andrew Baumann, Marcus Peinado, and Galen Hunt.
\newblock Shielding applications from an untrusted cloud with haven.
\newblock {\em ACM Transactions on Computer Systems (TOCS)}, 33(3):8, 2015.

\bibitem{benson2010network}
Theophilus Benson, Aditya Akella, and David~A Maltz.
\newblock Network traffic characteristics of data centers in the wild.
\newblock In {\em Proceedings of the 10th ACM SIGCOMM conference on Internet
  measurement}, pages 267--280. ACM, 2010.

\bibitem{brenner2016securekeeper}
Stefan Brenner, Colin Wulf, David Goltzsche, Nico Weichbrodt, Matthias Lorenz,
  Christof Fetzer, Peter Pietzuch, and R{\"u}diger Kapitza.
\newblock Securekeeper: Confidential zookeeper using intel sgx.
\newblock In {\em Proceedings of the 16th Annual Middleware Conference
  (Middleware)}, 2016.

\bibitem{cardellini1999dynamic}
Valeria Cardellini, Michele Colajanni, and Philip~S Yu.
\newblock Dynamic load balancing on web-server systems.
\newblock {\em IEEE Internet computing}, 3(3):28--39, 1999.

\bibitem{chen2008overshadow}
Xiaoxin Chen, Tal Garfinkel, E~Christopher Lewis, Pratap Subrahmanyam, Carl~A
  Waldspurger, Dan Boneh, Jeffrey Dwoskin, and Dan~RK Ports.
\newblock Overshadow: a virtualization-based approach to retrofitting
  protection in commodity operating systems.
\newblock In {\em ACM SIGARCH Computer Architecture News}, volume~36, pages
  2--13. ACM, 2008.

\bibitem{coughlin2017trusted}
Michael Coughlin, Eric Keller, and Eric Wustrow.
\newblock Trusted click: Overcoming security issues of nfv in the cloud.
\newblock In {\em Proceedings of the ACM International Workshop on Security in
  Software Defined Networks \& Network Function Virtualization}, pages 31--36.
  ACM, 2017.

\bibitem{criswell2014virtual}
John Criswell, Nathan Dautenhahn, and Vikram Adve.
\newblock Virtual ghost: Protecting applications from hostile operating
  systems.
\newblock {\em ACM SIGPLAN Notices}, 49(4):81--96, 2014.

\bibitem{dyer2001building}
Joan~G Dyer, Mark Lindemann, Ronald Perez, Reiner Sailer, Leendert Van~Doorn,
  and Sean~W Smith.
\newblock Building the ibm 4758 secure coprocessor.
\newblock {\em Computer}, 34(10):57--66, 2001.

\bibitem{tpm}
TC~Group et~al.
\newblock Trusted platform module library specification (tpm2. 0), 2013.

\bibitem{hofmann2013inktag}
Owen~S Hofmann, Sangman Kim, Alan~M Dunn, Michael~Z Lee, and Emmett Witchel.
\newblock Inktag: Secure applications on an untrusted operating system.
\newblock In {\em ACM SIGARCH Computer Architecture News}, volume~41, pages
  265--278. ACM, 2013.

\bibitem{sgx1}
Intel.
\newblock Intel software guard extensions developer guide, 2017.

\bibitem{sgx2}
Intel.
\newblock Intel software guard extensions sdk for linux os developer reference,
  2017.

\bibitem{dpdk}
DPDK Intel.
\newblock Data plane development kit, 2014.

\bibitem{johnson2016intel}
Simon Johnson, Vincent Scarlata, Carlos Rozas, Ernie Brickell, and Frank
  Mckeen.
\newblock Intel software guard extensions: Epid provisioning and attestation
  services.
\newblock {\em ser. Intel Corporation}, 2016.

\bibitem{kaplan2016amd}
David Kaplan, Jeremy Powell, and Tom Woller.
\newblock Amd memory encryption.
\newblock {\em White paper, Apr}, 2016.

\bibitem{kawashima2016host}
Ryota Kawashima, Shin Muramatsu, Hiroki Nakayama, Tsunemasa Hayashi, and
  Hiroshi Matsuo.
\newblock A host-based performance comparison of 40g nfv environments focusing
  on packet processing architectures and virtual switches.
\newblock In {\em Software-Defined Networks (EWSDN), 2016 Fifth European
  Workshop on}, pages 19--24. IEEE, 2016.

\bibitem{kawashima2017evaluation}
Ryota Kawashima, Hiroki Nakayama, Tsunemasa Hayashi, and Hiroshi Matsuo.
\newblock Evaluation of forwarding efficiency in nfv-nodes toward predictable
  service chain performance.
\newblock {\em IEEE Transactions on Network and Service Management}, 2017.

\bibitem{kent2005ip}
Stephen Kent.
\newblock Ip encapsulating security payload (esp).
\newblock 2005.

\bibitem{kim2017enhancing}
Seong~Min Kim, Juhyeng Han, Jaehyeong Ha, Taesoo Kim, and Dongsu Han.
\newblock Enhancing security and privacy of tor's ecosystem by using trusted
  execution environments.
\newblock In {\em NSDI}, pages 145--161, 2017.

\bibitem{kim2015first}
Seongmin Kim, Youjung Shin, Jaehyung Ha, Taesoo Kim, and Dongsu Han.
\newblock A first step towards leveraging commodity trusted execution
  environments for network applications.
\newblock In {\em Proceedings of the 14th ACM Workshop on Hot Topics in
  Networks}, page~7. ACM, 2015.

\bibitem{lan2016embark}
Chang Lan, Justine Sherry, Raluca~Ada Popa, Sylvia Ratnasamy, and Zhi Liu.
\newblock Embark: Securely outsourcing middleboxes to the cloud.
\newblock In {\em NSDI}, volume~16, pages 255--273, 2016.

\bibitem{lengyel2014multi}
Tamas~K Lengyel, Thomas Kittel, Jonas Pfoh, and Claudia Eckert.
\newblock Multi-tiered security architecture for arm via the virtualization and
  security extensions.
\newblock In {\em Database and Expert Systems Applications (DEXA), 2014 25th
  International Workshop on}, pages 308--312. IEEE, 2014.

\bibitem{leung2008possible}
Adrian Leung, Liqun Chen, and Chris Mitchell.
\newblock On a possible privacy flaw in direct anonymous attestation (daa).
\newblock {\em Trusted Computing-Challenges and Applications}, pages 179--190,
  2008.

\bibitem{li2014minibox}
Yanlin Li, Jonathan~M McCune, James Newsome, Adrian Perrig, Brandon Baker, and
  Will Drewry.
\newblock Minibox: A two-way sandbox for x86 native code.
\newblock In {\em USENIX Annual Technical Conference}, pages 409--420, 2014.

\bibitem{mckeen2013innovative}
Frank McKeen, Ilya Alexandrovich, Alex Berenzon, Carlos~V Rozas, Hisham Shafi,
  Vedvyas Shanbhogue, and Uday~R Savagaonkar.
\newblock Innovative instructions and software model for isolated execution.
\newblock {\em HASP@ ISCA}, 10, 2013.

\bibitem{morreale2015software}
Patricia~A Morreale and James~M Anderson.
\newblock {\em Software Defined Networking: Design and Deployment}.
\newblock CRC Press, 2015.

\bibitem{nadareishvili2016microservice}
Irakli Nadareishvili, Ronnie Mitra, Matt McLarty, and Mike Amundsen.
\newblock {\em Microservice Architecture: Aligning Principles, Practices, and
  Culture}.
\newblock " O'Reilly Media, Inc.", 2016.

\bibitem{orenbach2017eleos}
Meni Orenbach, Pavel Lifshits, Marina Minkin, and Mark Silberstein.
\newblock Eleos: Exitless os services for sgx enclaves.
\newblock In {\em EuroSys}, pages 238--253, 2017.

\bibitem{peinado2004ngscb}
Marcus Peinado, Yuqun Chen, Paul England, and John Manferdelli.
\newblock Ngscb: A trusted open system.
\newblock In {\em Australasian Conference on Information Security and Privacy},
  pages 86--97. Springer, 2004.

\bibitem{pinto2014towards}
Sandro Pinto, Daniel Oliveira, Jorge Pereira, Nuno Cardoso, Mongkol
  Ekpanyapong, Jorge Cabral, and Adriano Tavares.
\newblock Towards a lightweight embedded virtualization architecture exploiting
  arm trustzone.
\newblock In {\em Emerging Technology and Factory Automation (ETFA), 2014
  IEEE}, pages 1--4. IEEE, 2014.

\bibitem{pires2016secure}
Rafael Pires, Marcelo Pasin, Pascal Felber, and Christof Fetzer.
\newblock Secure content-based routing using intel software guard extensions.
\newblock In {\em Proceedings of the 17th International Middleware Conference},
  page~10. ACM, 2016.

\bibitem{randles2010comparative}
Martin Randles, David Lamb, and A~Taleb-Bendiab.
\newblock A comparative study into distributed load balancing algorithms for
  cloud computing.
\newblock In {\em Advanced Information Networking and Applications Workshops
  (WAINA), 2010 IEEE 24th International Conference on}, pages 551--556. IEEE,
  2010.

\bibitem{romanow2006media}
Allyn Romanow.
\newblock Media access control (mac) security.
\newblock {\em IEEE 802.1 AE}, 2006.

\bibitem{santos2009towards}
Nuno Santos, Krishna~P Gummadi, and Rodrigo Rodrigues.
\newblock Towards trusted cloud computing.
\newblock {\em HotCloud}, 9(9):3, 2009.

\bibitem{schuster2015vc3}
Felix Schuster, Manuel Costa, C{\'e}dric Fournet, Christos Gkantsidis, Marcus
  Peinado, Gloria Mainar-Ruiz, and Mark Russinovich.
\newblock Vc3: Trustworthy data analytics in the cloud using sgx.
\newblock In {\em Security and Privacy (SP), 2015 IEEE Symposium on}, pages
  38--54. IEEE, 2015.

\bibitem{seo2017sgx}
Jaebaek Seo, Byounyoung Lee, Seongmin Kim, Ming-Wei Shih, Insik Shin, Dongsu
  Han, and Taesoo Kim.
\newblock Sgx-shield: Enabling address space layout randomization for sgx
  programs.
\newblock In {\em Proceedings of the 2017 Annual Network and Distributed System
  Security Symposium (NDSS), San Diego, CA}, 2017.

\bibitem{sherry2012making}
Justine Sherry, Shaddi Hasan, Colin Scott, Arvind Krishnamurthy, Sylvia
  Ratnasamy, and Vyas Sekar.
\newblock Making middleboxes someone else's problem: network processing as a
  cloud service.
\newblock {\em ACM SIGCOMM Computer Communication Review}, 42(4):13--24, 2012.

\bibitem{sherry2015blindbox}
Justine Sherry, Chang Lan, Raluca~Ada Popa, and Sylvia Ratnasamy.
\newblock Blindbox: Deep packet inspection over encrypted traffic.
\newblock {\em ACM SIGCOMM Computer Communication Review}, 45(4):213--226,
  2015.

\bibitem{shih2016s}
Ming-Wei Shih, Mohan Kumar, Taesoo Kim, and Ada Gavrilovska.
\newblock S-nfv: securing nfv states by using sgx.
\newblock In {\em Proceedings of the 2016 ACM International Workshop on
  Security in Software Defined Networks \& Network Function Virtualization},
  pages 45--48. ACM, 2016.

\bibitem{smith1999building}
Sean~W Smith and Steve Weingart.
\newblock Building a high-performance, programmable secure coprocessor.
\newblock {\em Computer Networks}, 31(8):831--860, 1999.

\bibitem{strackx2015idea}
Raoul Strackx, Pieter Philippaerts, and Fr{\'e}d{\'e}ric Vogels.
\newblock Idea: Towards an inverted cloud.
\newblock In {\em International Symposium on Engineering Secure Software and
  Systems}, pages 111--118. Springer, 2015.

\bibitem{ta2006splitting}
Richard Ta-Min, Lionel Litty, and David Lie.
\newblock Splitting interfaces: Making trust between applications and operating
  systems configurable.
\newblock In {\em Proceedings of the 7th symposium on Operating systems design
  and implementation}, pages 279--292. USENIX Association, 2006.

\bibitem{trach2017slick}
Bohdan Trach, Alfred Krohmer, Sergei Arnautov, Franz Gregor, Pramod Bhatotia,
  and Christof Fetzer.
\newblock Slick: Secure middleboxes using shielded execution.
\newblock {\em arXiv preprint arXiv:1709.04226}, 2017.

\bibitem{weichbrodt2016asyncshock}
Nico Weichbrodt, Anil Kurmus, Peter Pietzuch, and R{\"u}diger Kapitza.
\newblock Asyncshock: Exploiting synchronisation bugs in intel sgx enclaves.
\newblock In {\em European Symposium on Research in Computer Security}, pages
  440--457. Springer, 2016.

\end{thebibliography}

\end{document}